\documentclass[aps,prd,nofootinbib,superscriptaddress]{revtex4}
\usepackage{graphics}
\usepackage{epsfig}
\usepackage[english]{babel}
\usepackage{amsmath} 
\usepackage{verbatim} 
\usepackage{mathrsfs}
\usepackage{amsfonts}
\usepackage{amssymb}
\usepackage[latin1]{inputenc}
\usepackage{hyperref}

\newcommand{\beq}{\begin{equation}}
\newcommand{\eeq}{\end{equation}}
\newcommand{\nk}{\textbf{k}}
\newcommand{\dphi}{\delta \phi}
\newcommand{\x}{\textbf{x}}

\newcommand{\bra}{\langle}
\newcommand{\ket}{\rangle}
\newcommand{\mH}{\mathcal{H}}
\newcommand{\mP}{\mathcal{P}}
\newcommand{\mR}{\mathcal{R}}
\newcommand{\mW}{\mathcal{W}}
\newcommand{\barr}{\begin{eqnarray}}
\newcommand{\earr}{\end{eqnarray}}
\newcommand{\bea}{\begin{eqnarray*}}
\newcommand{\eea}{\end{eqnarray*}}
\newcommand{\nn}{\nonumber \\}
  \newcommand{\bepsilon}{\overline{\epsilon}}
  \newcommand{\dvarphi}{\delta \varphi}
  \newcommand{\bdelta}{\overline{\delta^2}}
  \newcommand{\eq}[1]{Eq. \eqref{#1}}
 \newcommand{\eqs}[1]{Eqs. \eqref{#1}}

\begin{document}

\title{Quasi-matter bounce and inflation in the light of the CSL model}

\author{Gabriel Le\'{o}n}
\email{gleon@df.uba.ar} \affiliation{Departamento de
F\'\i sica, Facultad de Ciencias Exactas y Naturales, Universidad
de Buenos Aires, Ciudad Universitaria - Pab.I, 1428 Buenos Aires,
Argentina}
\author{Gabriel R. Bengochea}
\email{gabriel@iafe.uba.ar} \affiliation{Instituto de Astronom\'\i
a y F\'\i sica del Espacio (IAFE), CONICET - Universidad de Buenos Aires, 1428 Buenos Aires, Argentina}
\author{Susana J. Landau}
\email{slandau@df.uba.ar} \affiliation{Departamento de
F\'\i sica, Facultad de Ciencias Exactas y Naturales, Universidad
de Buenos Aires and IFIBA, CONICET, Ciudad Universitaria - Pab.I, 1428 Buenos Aires,
Argentina}

\begin{abstract}

The Continuous Spontaneous Localization (CSL) model has been proposed as a possible
solution to the quantum measurement problem by modifying the Schr\"{o}dinger equation.
In this work, we apply the CSL model to two cosmological models of the early Universe: the matter
bounce scenario and slow roll inflation. In particular, we focus on the generation of the
classical primordial inhomogeneities and anisotropies that arise from the dynamical
evolution, provided by the CSL mechanism, of the quantum state associated to the quantum
fields. In each case, we obtained a prediction for the shape and the parameters
characterizing the primordial spectra (scalar and tensor), i.e. the amplitude, the
spectral index and the tensor-to-scalar ratio. We found that there exist CSL parameter values,
allowed by other non-cosmological experiments, for which our predictions
for the angular power spectrum of the CMB temperature anisotropy are consistent with the best fit
canonical model to the latest data released by the Planck Collaboration.

\end{abstract}

\maketitle

\section{Introduction}
\label{intro}

After approximately three decades since the cosmological inflationary paradigm was
conceived \cite{guth,starobinsky,linde,albrecht}, all of its generic predictions have
withstood the confrontation with
observational data, in particular, those coming from the Cosmic Microwave
Background (CMB) radiation \cite{planckcmb2,Planckinflation15,PlanckBicep15}. That has led a large
group of cosmologists to consider inflation as a well established theory of the early
Universe. Inflation was originally proposed to provide a solution to the
puzzles of the hot Big Bang theory (e.g. the horizon and flatness problems).
However, the modern success of inflation
is that, allegedly, it can
offer us an explanation about the origin of the primordial inhomogeneities
\cite{mukhanov81,starobinsky2,hawking,bardeen2}. The standard argument is also rather
pictorial: the quantum fluctuations of the vacuum associated to the inflaton are stretched
out to cosmological scales due to the accelerated expansion of the spacetime; those
fluctuations are considered the seeds of all large scale structures observed in the
Universe. Furthermore, in Ref. \cite{Martin2016} is investigated the detectability of possible
traces of the quantum nature regarding the primordial perturbations.

On the other hand, proponents of alternative scenarios to inflation argue that even if
it is the most fashionable model of the early Universe, that does not mean it is
necessarily true \cite{ijjas2,petercritics}. Furthermore, another feature that would make
alternative models worthwhile of study is that they might avoid some long
standing puzzles of the inflationary paradigm. Among those issues, we
can mention: the subject of eternal
inflation, a feature that is present in almost every model of inflation \cite{eternal}
and which also leads to the controversial topic of the multiverse; the initial
singularity problem and the \emph{trans-Planckian problem} for primordial perturbations
\cite{TPproblem}, which are related by the fact that one is interpolating the
solutions provided by General Relativity in regimes where it may no longer be valid;
and finally, it has been argued that the potentials associated to the inflaton, that best fit
the latest observed data, need to be fine-tuned \cite{ijjas,steinhardtplanck2015}.
Although the aforementioned problems are not considered real problems by some scientists
\cite{linde3,guth3}, others seem to disagree \cite{bbergerreview,steinhardtplanck2015}.
However, we think that if other alternative models can reproduce the
main results linked to inflation, we should make use of the observational data available
to test them.

One of the alternative models to inflation that seems to be consistent with the latest
data is the so called \emph{matter bounce scenario} (MBS)
\cite{bbergerreview,elizalde,jaume,amoros,haro,haro2,Cai2014,Cai2016}. In this cosmological model, the
initial singularity of the standard model is replaced by a non-singular bounce. That is,
instead of an ever-expanding Universe, it assumes an early contracting matter-dominated
Universe, which continues to evolve towards a bouncing phase and, later, enters into the
expanding-phase of standard cosmology. The Universe described by the MBS relies on a
single scalar field satisfying an equation of state that mimics that of a dust-like fluid.
Additionally, in order to describe successfully a bouncing phase with a single scalar
field, one needs to use cosmologies beyond the realm of General Relativity, such as, loop
quantum cosmologies, teleparallel $F(T)$ gravity or $F(R)$ gravity. Proponents of the MBS
claim that the potential associated to the scalar field is less fine-tuned than that of
inflation, and also solves the historical problem of requiring very special initial
conditions for the Big Bang model \cite{elizalde,jaume}, which originally motivated the
development of the inflationary framework. However, the MBS is not exactly problem free. A
complete assessment of the present conceptual issues is provided in Ref. \cite{jaume}. In
spite of not being completely finished from a theoretical point of view, the MBS is quite
simple in its treatment of the primordial perturbations. That makes it an interesting
case of study for the purpose of this article. In particular, the generation of the primordial perturbations is
depicted during the contracting phase, i.e. in a regime where gravity is well described
by General Relativity, and the perturbations correspond to inhomogeneities of a single
scalar field.

In addition to the prior puzzles and successes of inflation and the MBS, there remains an
important question: what is the precise physical mechanism that converts quantum
fluctuations of the vacuum into classical perturbations of the spacetime? This question has been
the subject of numerous works in the past and the consensus seems
to favor the decoherence framework
\cite{kiefer,halliwell,kiefer2,polarski,grishchuk}.\footnote{Although for the reasons
exposed in Refs. \cite{shortcomings,susana2013} we do not find such posture satisfactory.}
Nevertheless, decoherence cannot address that question by itself.\footnote{See comments by
Mukhanov on pages 347-348 of Ref. \cite{mukhanovbook} and by Weinberg on page 476 of
Ref. \cite{weinberg2008}. } In other words, even if one would choose (or not) to embrace
the
decoherence program, a particular interpretation of Quantum Mechanics must be selected
(implicitly or explicitly). The Copenhagen--orthodox--interpretation requires to identify
a notion of observer that performs a measurement on the system; which, in the decoherence
framework, is equivalent to identify the unobservables or external degrees of freedom of
the system. It is not clear how to do such identifications if the system is the early Universe. Other interpretations such as
many-worlds, consistent histories and hidden variables formulations, might be adopted with
varying degrees of success (see for instance \cite{pintoneto,valentini,goldstein}).

In the present article, in order to address the quantum-to-classical transition of
the primordial perturbations, we will choose to work with the Continuous
Spontaneous Localization (CSL) model. The CSL model belongs to a large class of models
known as objective dynamical reduction models or simply called collapse models.
Collapse models attempt to provide a solution to the measurement
problem of Quantum Mechanics \cite{GRW,pearle,bassi,bassi2012,bassi2012b}. The proponents
of these models state that the measurement problem originates from the linear character of
the quantum dynamics encoded in the Schr\"odinger equation. The common idea shared in
these collapse models is to introduce some nonlinear
stochastic corrections to the Schr\"odinger equation that breaks its linearity. According
to the collapse models, a noise field couples nonlinearly with the system (usually with
the spatial degree of freedom of the system), inducing a spontaneous random
localization of the wave function in a sufficiently small region of the space. Suitably
chosen collapse parameters make sure that micro-systems evolve essentially (but not
exactly) following the dynamics provided by the Schr\"odinger equation, while
macro-systems are extremely sensible to the nonlinear effects resulting in a perfectly
localization of the wave function. Furthermore, there is no need to mention or to
introduce a notion of an observer/measurement device as in the
Copenhagen interpretation, which is a desired feature in the context of the early
Universe and cosmology in general.

The CSL model has been applied before to the inflationary Universe in an attempt
to explain the quantum-to-classical transition of the primordial perturbations
\cite{jmartin,pedro,hinduesS,hinduesT,LB15}. Also, recently a new effective collapse mechanism,
independent of the CSL model, has been proposed to deal with the measurement problem
during the inflationary era \cite{Alexander2016}. However, among those works, the key role
played by the collapse mechanism varies and also yields different predictions for the
primordial power spectrum, some of which might be consistent with the observational data.
In the present article, we will subscribe to the conceptual point of view first presented
in \cite{PSS,pedro}, which was developed within the semiclassical gravity framework, and
latter in \cite{gabriel,LB15} was extended to the standard quantization procedure of the
primordial perturbations using the so called Mukhanov-Sasaki variable
\cite{mukhanov81,sasaki}. The main role that we advocate for the
dynamical reduction mechanism of the state vector, modeled in this paper by the CSL
model, is to directly generate the primordial curvature perturbations.
Specifically, the initial state of the quantum field--the vacuum state--evolves
dynamically according to the modified Schr\"{o}dinger equation provided by the CSL model.
This evolution leads to a final state that does not share the initial symmetry of the
vacuum, i.e. it is not homogeneous and isotropic.\footnote{For a formal proof of this
statement see   Appendix A of Ref. \cite{susana2013} and Appendix A of Ref. \cite{LB15}.
}
In this way, the collapse mechanism generates the inhomogeneities and anisotropies of the
matter fields. These asymmetries are codified in the evolved quantum state and, thus,
are responsible for generating the
perturbations of the spacetime.\footnote{We encourage the reader to consult Refs.
\cite{PSS,susana2013,LB15} for a detailed exposition of the concepts
involved in our approach and its relation with the collapse of the wave function.}

Note that the previous prescription, regarding our approach to address the birth of the
primordial perturbations, does not require the inclusion of an exponential expansion
phase in the Universe that ``stretches out'' the quantum fluctuations of the vacuum (or
the squeezing of the field variables as usually argued).  Therefore, in principle, it
should be possible to extend our picture to alternative scenarios dealing with the origin
of the cosmological perturbations. Moreover, since the cosmic observations are well
constrained, it should also be feasible to test the predictions that result from
applying our framework in those alternative
cosmological models. In the present article, we focus on the implementation of the CSL
model within the framework of the MBS and, in parallel, we present the same appliance of
the CSL model to the slow roll inflationary model of the early Universe. In this way, we
can appreciate more clearly where the CSL model enters into the picture; particularly at
the moment when computing the theoretical predictions. The main motivation behind the
present work is that if the CSL model can be truly considered as a physical model of the
quantum world, which also avoids the standard quantum measurement problem, then it should
also be possible to use it in different contexts from the traditional laboratory
settings. The cosmological context provides a rich avenue to explore such foundational
issues and, more important, there exist sufficient precise data to test the initial
hypotheses. As a consequence, we will analyze the predictions resulting from implementing
the CSL model in the MBS and in the inflationary model of the early Universe, and we will
compare the corresponding results with the one provided by the best fit standard
cosmological model.
Additionally, we will focus on the range of values allowed for the parameters of the CSL
model, experimentally tested \cite{bassibounds2016,bassibounds22016} in
non-cosmological frameworks.

The paper is organized as follows: in Sect. \ref{CSLreview}, we provide a very brief
synopsis of the main features of the CSL model, with particular emphasis on those that
will be useful for the next sections. In Sect. \ref{reviewMBSinflation}, we present the
characterization of the primordial perturbations within the two cosmological models that
we are considering, i.e. the MBS and standard slow roll inflation. In Sect. \ref{connection},
we show the connection between the observational quantities and the theoretical predictions that result from adopting our
conceptual point of view concerning the CSL model. In Sect. \ref{SecImplementation}, we
explicitly show the implementation of the CSL model to the MBS
and inflation, and we also present the predictions for the primordial power spectra (scalar and tensor) in each
case. In Sect. \ref{discussionPS}, we discuss the implications of the results obtained;
additionally, we compare the predicted scalar power spectra with the standard one.
In Sect. \ref{effectsCMB}, we analyze the viability of the CSL model using the data
extracted from the CMB when considering the best fit cosmological model.
Finally, in Sect. \ref{conclusions}, we end with our conclusions. We include an Appendix containing
the computational details that led to the results presented in Sect.
\ref{SecImplementation}.

\section{A concise synopsis of the CSL model}
\label{CSLreview}

In this section, we provide a brief summary of the relevant features of the CSL model;
for a detailed review, we refer the reader to Refs. \cite{bassi,bassi2012}.

In the CSL model, the modification of the Schr\"odinger equation
induces a collapse of the wave function towards one of the possible eigenstates of an
operator $\hat \varTheta$, called the collapse operator, with certain rate $\lambda$. The
self-induced collapse is due to the interaction of the system with a background noise
$\mW(t)$ that can be considered as a continuous-time stochastic process of the Wiener
kind.  The modified Schr\"{o}dinger equation drives the time evolution of an initial
state
as
\beq\label{csl}
| \Psi, t \ket = \hat T \exp \left\{-\int_{t_0}^t dt' \left[i \hat H + \frac{1}{4\lambda}
\left(\mW(t')-2\lambda \hat \varTheta \right)^2 \right] \right\} |\Psi, t_0 \ket,
\eeq
with $\hat T$ the time-ordering operator. The probability associated with a
particular realization of $\mW(t)$ is,
\beq\label{prob}
P[\mW(t),t] D\mW(t) = \bra \Psi, t | \Psi, t \ket \prod_{t_i=t_0}^t
\frac{d\mW(t_i)}{\sqrt{2\pi \lambda/dt}}.
\eeq
The norm of the state $|\Psi, t\ket$ evolves dynamically, and Eq. \eqref{prob}
implies that the most probable state will be the one with the largest norm.
From Eqs. \eqref{csl} and \eqref{prob}, it can be derived the evolution equation
of the density matrix operator $\hat \rho$. That is,
\beq\label{rhodt}
\frac{d \hat \rho}{dt} = -i [\hat H, \hat \rho] - \frac{\lambda}{2} [\hat
\varTheta, [\hat \varTheta, \hat \rho]].
\eeq
The density matrix operator can be used to obtain the ensemble average of the
expectation value of an operator $\overline{\bra \hat O \ket}=\textrm{Tr}[\hat
O
\hat \rho]$. Henceforth, from Eq. \eqref{rhodt} it follows that
\beq\label{operadorcsl}
\frac{d}{dt} \overline{ \bra \hat O \ket } = -i \overline{\langle [\hat O, \hat H]\rangle } -
\frac{\lambda}{2} \overline{\langle [\hat \varTheta, [\hat \varTheta, \hat O]]\rangle }.
\eeq
The average is over possible realizations of the noise $\mW(t)$, each realization
corresponding to a single outcome of the final state $| \Psi,t \ket$.

One of the most important features of collapse models is the so-called amplification
mechanism. That is, assuming that the reduction (collapse) rates for the $M$
constituents of a macroscopic object are equal ($\lambda^i = \lambda$), it can be proved
that the reduction rate for the center of mass of an $M$-particle system is amplified by
a factor of $M$ with respect to that of a single constituent \cite{GRW,GPR}; in other
words, $\lambda_{\textrm{macro}}= M\lambda$.

The parameter $\lambda$ sets the strength of the collapse process. In the original model,
proposed by Ghirardi-Rimmini-Webber (GRW), the authors suggested a value of
$\lambda_{\textrm{GRW}} \simeq 10^{-16}$ s$^{-1}$ for $r_C \simeq 100$ nm. However, Adler
suggested a greater value $\lambda_{\textrm{Adler}} \simeq 10^{-8}$ s$^{-1}$ for $r_C
\simeq 100$ nm \cite{adler2} (the parameter $r_C$ is called the correlation length of the
noise and provides a measure for the spatial resolution of the collapse
\cite{GRW,GPR,bassi}). Recent experiments have been devised  to set
bounds on the parameter $\lambda$ \cite{mohammed,mohammed2}. Furthermore, it is claimed
that matter-wave interferometry provides the most generic way to experimentally test the
collapse models  \cite{bassibounds2016,bassibounds22016}. Those results suggest that the
range between $\lambda_{\textrm{GRW}}$ and $\lambda_{\textrm{Adler}}$ is still viable for
some variations of the original CSL model (e.g. by considering non-white noise).

Henceforth, the main characteristics of the
CSL model are: (1) The modification to the Schr\"odinger equation is nonlinear and leads
to a breakdown of the superposition principle for macroscopic objects; (2) The random nature of
Quantum Mechanics is concealed in the noise $\mW(t)$ and is consistent with Born's rule;
(3) An amplification mechanism exists, through the parameter $\lambda$ which is related to
the strength of the collapse. This strength is weak for microscopic objects and strong for
macroscopic bodies.

Another main aspect of the collapse models is that the collapse mechanism injects energy
into the system. In fact, previous works have performed a preliminary analysis using
cosmological data to set bounds on the value of $\lambda$ \cite{kinjalk}. The energy
increase is minimal, e.g. for a particle of mass $m=10^{-23}$ g, one obtains $\delta E/t
\simeq 10^{-25}$ eV s$^{-1}$ \cite{bassi}. In other words, an increase of $10^{-8}$ eV
will take $10^{10}$ years. However, even if the energy increase can be ignored at the
phenomenological level, a more realistic model should remove this issue.

Moreover, the increase of energy in the collapse models leads to difficulties
when trying to formulate relativistic collapse models. Additionally, the collapse
mechanism
occurs in such a way that is nonlocal. This implies that the collapse of the wave
function must be instantaneous or superluminal (but the nonlocal features cannot be
exploited to send signals at superluminal speed). Also, the nonlocality is necessary
to ensure that the models are consistent with the violation of Bell's inequalities.
Several relativistic models have been proposed so far  \cite{tumulka,bedingham,pearle3},
none of which can be considered completely finished. In spite of the lack of a
relativistic collapse model, we will apply the CSL model to the primordial Universe,
i.e. to inflation and the MBS, but in order to provide a
more detailed picture, we need first to establish the mathematical framework of the
primordial Universe in the two approaches considered in this work.

\section{Two approaches: accelerated expansion or quasi-matter contraction}\label{reviewMBSinflation}

This section presents the details of the two cosmological approaches, describing the dynamics
of the Universe, that we will be considering in the rest of the manuscript. In particular, we
are going to work with the following two scenarios:

\begin{enumerate}
 \item An accelerated expansion of the early Universe given by the simplest inflationary
model, that is, a single scalar field in the slow roll approximation with canonical
kinetic term. Since such a model is probably very well known for most readers, we
will not dwell into much detail here.

 \item The MBS  \cite{elizalde,jaume,amoros,haro,haro2}, a cosmological model in
which the Universe undertakes a
 quasi-matter contracting phase, then experiences a non-singular bounce and finally
enters into the standard cosmological expansion. Since in this model the primordial
perturbations are born during the contracting stage of the Universe, we will focus
exclusively on that cosmic stage. We will refer to such a stage as the quasi-matter
contracting Universe (QMCU).
\end{enumerate}

\subsection{The background}

The inflationary Universe and the QMCU are both described by Einstein equations $G_{ab} =
8 \pi GT_{ab}$ ($c=1$), while the matter fields are characterized by a single scalar
field. In the case of inflation the scalar field is the inflaton $\phi$, and in the QMCU
the scalar field will be denoted by $\varphi$.

As mentioned earlier, for inflation, we will consider standard slow roll inflation. In
that case, the background spacetime is
described by a quasi-de Sitter Universe, characterized by $\mH
\simeq -1/[\eta(1-\epsilon)]$, with $\mH \equiv a'/a$ the conformal expansion
rate, $a$ being the scale factor and the slow roll parameter is defined as
$\epsilon \equiv 1-\mH'/\mH^2$; a prime denotes partial derivative with respect
to conformal time $\eta$. The energy density of the Universe is dominated by
the potential of the inflaton $V$, and during slow roll inflation
the condition $ \epsilon \simeq  M_P^2/2 (\partial_\phi V/V)^2 \ll 1$ is satisfied, with
$M_P^2 \equiv (8\pi G)^{-1}$ the reduced Planck mass. Since we will work in a
full quasi-de Sitter expansion, another useful parameter to characterize slow roll
inflation is the second slow roll parameter, i.e.
$\delta \equiv \epsilon - \epsilon'/2\mH \epsilon \ll 1$.

In the case of the QMCU, the starting point is also a flat FLRW geometry that leads to the
Friedmann and conservation equations. The field $\varphi$ is separated into an
homogeneous part $\varphi_0(\eta)$  plus small inhomogeneities $\dvarphi(\x,\eta)$. The
homogeneous part satisfies
\beq
\mH^2 = \frac{a^2}{3 M_P} \left( \frac{\varphi_0'^2}{2a^2} + W \right); \quad
\varphi_0''+
2\mH \varphi_0'+a^2 \partial_\varphi W =0,
\eeq
where $W$ is the potential associated to the field $\varphi$.

In the QMCU, it is assumed that the equation of state associated to the scalar field
almost mimics that of ordinary matter, i.e. $P = \omega \rho$ such that $|\omega| \ll 1$;
the latter
implies $\varphi_0'^2 \simeq 2 a^2 W$. Consequently, the scale factor (in conformal
time) evolves as $a(\eta) \simeq \eta^2/9$.

The quasi-matter contraction is characterized with a small parameter $|\bepsilon| \ll 1$,
which plays the same role as the slow roll parameter in inflation. The parameter
$\bepsilon$ is defined as (see e.g. \cite{elizalde})
\beq
\bepsilon \equiv - \frac{2}{3} \left( \frac{1}{2} + \frac{\mH'}{\mH^2} \right) \simeq
\frac{1}{3} \left( \frac{\partial_\varphi W}{W}  \right)^2 -1.
\eeq
The case $\bepsilon = 0$ corresponds to an exact matter-dominated contracting phase
(note that $\bepsilon = \omega$). Furthermore, for sake of completeness we
introduce another parameter
\beq
\bdelta \equiv \frac{\bepsilon'}{2\mH} \simeq - \partial_\varphi \left(
\frac{\partial_\varphi W}{W}   \right),
\eeq
such that $|\bdelta| \ll |\bepsilon|$. The parameter
$|\bdelta|$ is analog to the
$\delta$ parameter of slow roll inflation and it is related to the running of the
spectral index in the QMCU model (see e.g. \cite{elizalde}).

As is well known, it is not straightforward to accomplish a
non-singular bounce within the framework of General Relativity by considering a single canonical scalar
field, since the null energy condition (NEC) is violated (see for instance \cite{petercritics,jaume}). As a consequence, one possible option is to work with cosmologies within the context of modified gravity theories. In the case of the QMCU presented in
\cite{elizalde,jaume}, the authors worked within the framework of holonomy corrected loop quantum cosmology and
teleparallel $F(T)$ gravity.

It is also important to note that even if a non-singular bounce cannot be achieved within
General Relativity, the origin of the primordial perturbations is assumed to take place
during the contracting (pre-bounce) phase of the Universe, where the curvature and
energy scales are low enough to be described by General Relativity. On the other hand,
one must present
the conditions that need to be fulfilled such that  the shape of the primordial
spectrum, associated to the perturbations, remains practically unchanged when passing
through the bounce. We will discuss this subject in more detail in the next
section.

\subsection{Perturbations}

In the inflationary Universe and in the QMCU, one can separate the scalar field into an
homogeneous part plus small inhomogeneous perturbations. Moreover, the metric associated
to the spacetime, in both cases, is described by a FLRW background metric plus
perturbations; which are classified as scalar,
vector and tensor types (in this paper we will not consider vector
perturbations).  One useful quantity to describe the scalar (an also the tensor)
perturbations is the so called Mukhanov-Sasaki (MS) variable. During inflation, the MS
variable is defined by
\beq\label{vMS}
v(\x,\eta) \equiv a(\eta) \left[ \dphi(\x,\eta) + \frac{\phi_0'(\eta)}{\mH(\eta)}
\Phi(\x,\eta) \right],
\eeq
with $\Phi$ the gauge invariant quantity known as the Bardeen potential \cite{Bardeen},
which, in the longitudinal gauge, corresponds to the curvature perturbation.  A similar
expression to Eq. \eqref{vMS} can be used in the QMCU by replacing the fields
$\phi_0'$ and $\dphi$ with $\varphi_0'$ and $\dvarphi$, respectively. The advantage of
relying on the MS variable is that, when
expanding the action of a scalar field minimally coupled to gravity
into second order scalar perturbations, one obtains $\delta^{(2)} S = \frac{1}{2}
\int d\eta d^3 \nk \mathcal{L}$, where
\beq\label{accionv}
\mathcal{L} =  v_{\nk}' v_{\nk}^{\star '} - k^2 v_{\nk} v_{\nk}^{\star}
-\frac{z'}{z}  \left(  v_{\nk} v_{\nk}^{\star '}  +  v_{\nk}'  v_{\nk}^{\star}
\right) +  \left( \frac{z'}{z} \right)^2 v_{\nk} v_{\nk}^{\star},
\eeq
with $v_{\nk}$ the Fourier modes associated to the MS variable, $z = a\phi_0'/\mH$
during inflation, and $z =a\varphi_0'/\mH$ when considering the QMCU. However, it is
important to note that during the bouncing phase, the action given by the Lagrangian in
Eq. \eqref{accionv}
remains the same but the expression for $z$ changes (see Ref. \cite{haro} for an
explicit calculation within $F(T)$ theories). On the other hand, during the contraction
phase, the quantity $z'/z$ can be written explicitly in terms of the QMCU parameter
$\bar \epsilon$, as in a similar fashion using the
slow roll inflation parameters, i.e.
\begin{eqnarray}\label{zbeta}
  \frac{z'(\eta)}{z(\eta)} = \frac{\beta}{\eta};  \qquad \textrm{where} \qquad \beta
\equiv \begin{cases} -(1+2\epsilon-\delta)
\:\: \textrm{if assuming the inflationary universe }\\
2(1-3\bar \epsilon ) \:\: \textrm{if assuming the QMCU }  \\
\end{cases}
\end{eqnarray}
Note that, since $|\bdelta| \ll |\bepsilon|$, the $\bdelta$ parameter does not
enter into the expression $z'/z$ at first order for the QMCU.

The CSL model is based on a nonlinear modification to the Schr\"odinger equation;
consequently,
it will be advantageous to perform the quantization of the perturbations in the
Schr\"odinger picture, where the relevant physical objects are the Hamiltonian and the
wave functional. The Hamiltonian associated to $\mathcal{L}$ in Eq. \eqref{accionv} is $H=
\frac{1}{2} \int d^3 \nk_{}\: (H_{\nk}^R +H_{\nk}^I)$, with
\beq\label{ham}
H^{R,I}_{\nk} = p_{\nk}^{R,I} p_{\nk}^{R,I} + \frac{z'}{z} \left(
v_{\nk}^{R,I} p_{\nk}^{R,I} + v_{\nk}^{R,I} p_{\nk}^{R,I} \right)
+ k^2 v_{\nk}^{R,I} v_{\nk}^{R,I},
\eeq
where the indexes $R,I$ denote the real and imaginary parts of $v_{\nk}$ and
$p_{\nk}$. The canonical conjugated momentum associated to $v_{\nk}$ is $p_{\nk} =
\partial
\mathcal{L}/ \partial v_{\nk}^{\star '}$, i.e.
\beq\label{momento}
p_{\nk} =  v_{\nk}'-\frac{z'}{z} v_{\nk}.
\eeq
Since $v(\x,\eta)$ is a real field, $v_{\nk}^\star = v_{-\nk}$.

We promote $v_{\nk}$ and $p_{\nk}$ to quantum operators, by
imposing canonical commutations relations $[\hat{v}_{\nk}^{R,I},
\hat{p}_{\nk'}^{R,I}] = i\delta (\nk-\nk')$.

In the Schr\"{o}dinger picture, the wave functional $\Psi[v(\x,\eta)]$
characterizes the state of the system. Furthermore, in Fourier space, the wave functional
can be factorized into modes components $\Psi[v(\x,\eta)] = \Pi_{\nk}
\Psi_{\nk}^R (v_{\nk}^R)\Psi_{\nk}^I (v_{\nk}^I)$. From now on, we will deal
with each mode separately.  Henceforth, each mode of the wave functional,
associated to the real and imaginary parts of the canonical variables,
satisfies the Schr\"{o}dinger equation $\hat H_{\nk}^{R,I} \Psi_{\nk}^{R,I} = i
\partial\Psi_{\nk}^{R,I}/\partial \eta$, with the Hamiltonian provided by
\eqref{ham}. Note that one can also choose to work with the wave functional in the
momentum representation, i.e.  $\Psi[p(\x,\eta)] = \Pi_{\nk} \Psi_{\nk}^R (p_{\nk}^R)
\Psi_{\nk}^I (p_{\nk}^I)$.

The standard assumption is that, at an early conformal time $\tau \to -\infty$, the modes
are in their adiabatic ground state, which is a Gaussian centered at zero with certain
spread. This applies to both, the inflationary Universe and the QMCU. In addition, this
ground state is commonly referred to as the Bunch-Davies vacuum. Thus, the conformal time
$\eta$ is in the range $[\tau,0^{-})$.

Given that the initial quantum state is Gaussian, its shape will be  preserved during
its evolution. The explicit expression of the Gaussian state, in the field
representation, is:
\beq\label{psiondav}
\Psi^{R,I}(\eta,v_{\nk}^{R,I}) = \exp\left[- A_{k}(\eta)(v_{\nk}^{R,I})^2 +
 B_{k}(\eta)v_{\nk}^{R,I} +  C_{k}(\eta)\right],
\eeq
and, equivalently, in the momentum representation
\beq\label{psiondap}
\Psi^{R,I}(\eta,p_{\nk}^{R,I}) = \exp\left[-\tilde A_{k}(\eta)(p_{\nk}^{R,I})^2 +
\tilde B_{k}(\eta)p_{\nk}^{R,I} + \tilde C_{k}(\eta)\right].
\eeq

Therefore, the wave functional evolves according to Schr\"{o}dinger equation, with initial
conditions given by $A_k (\tau ) = k/2,  {\tilde{A}}_k (\tau )= 1/2k, B_k (\tau )={\tilde{B}}_k(\tau ) = C_k(\tau )=
{\tilde{C}}_k(\tau ) = 0 $ corresponding to the Bunch-Davies
vacuum, which is perfectly homogeneous and isotropic in the sense of a vacuum
state in quantum field theory. The fact that we are introducing the wave functional in
the field and momentum representations is related to the choice of the collapse operator
in the CSL model, i.e., since there is no physical reason to choose one
over the other, both choices are equally acceptable (at least from the phenomenological
point of view). In the next section, we will show how to extract the physical quantities
from the theory to be compared with the observations.

\section{Theoretical predictions and observational
quantities}\label{connection}

We begin this section by making some key remarks about the conceptual aspects of our
approach and, then, we proceed to identify the relevant physical quantities that will be
related with the observed data. We encourage the reader to consult Refs.
\cite{PSS,shortcomings,susana2013} for a complete discussion regarding our full picture
of the role played by the dynamical reduction of wave function in the cosmological
setting. As a matter of fact, the relation between the observables and the
predictions from the theory, using the Mukhanov-Sasaki variable during inflation and the CSL model,
has been previously exposed in \cite{LB15}; however, in this section we reproduce the key
arguments of such a reference to make the present paper as self-contained as possible.
Thus, there is no original work in the following of this section.

The main role for invoking the collapse of the wave function is to find a physical
mechanism for breaking the initial homogeneity and isotropy associated to both, the
quantum state and the spacetime. More specifically, we assume
that a nonlinear modification to the Schr\"odinger equation, which in the present work
is provided by the CSL model, can break the homogeneity and isotropy associated to the
vacuum state and, in turn, it can generate the metric perturbations, which correspond to the
primordial curvature perturbation.

Note that in the literature one can found statements suggesting that the vacuum
fluctuations somehow become classical when the proper wavelength associated to the
perturbations becomes larger than the Hubble radius \cite{kiefer,riotto}. Nevertheless,
there is nothing in the dynamics governed by the traditional Schr\"odinger equation that
can change the symmetry of the vacuum state, the symmetry being the homogeneity and
isotropy. As a consequence, if the quantum state is perfectly symmetric and the Quantum
Theory teaches us that the symmetries of a physical system must be encoded in the quantum
state, then there is no clear way to describe the inhomogeneities and anisotropies of the
spacetime in the quantum sense. If the quantum state of the system is perfectly
symmetric, then its classical description must also be exactly symmetric. Thus,
there is a lack of a proper explanation concerning the emergence of the
primordial inhomogeneities and anisotropies in the Universe.
That is why some non-standard interpretations of Quantum Mechanics, that make use of the
Schr\"odinger equation (e.g. many-worlds, consistent histories, etc.), cannot provide a
satisfactory answer to the problem at hand. It is important to note that the previous
discussion applies to both cosmological models, the QMCU and inflation.

The modified Schr\"odinger equation given by
the CSL model can successfully change the symmetries of the vacuum state and, at the same
time, be responsible for the birth of the primordial curvature perturbation.

Specifically,  in the comoving gauge, the curvature perturbation $\mR(\x,\eta)$ and the
MS variable $v(\x,\eta)$ are related by $\mR(\x,\eta) = v(\x,\eta)/z(\eta)$. Thus, a
quantization of $ v(\x,\eta)$ implies a quantization of $ \mR(\x,\eta)$. The
question that arises now is: how to relate the quantum objects $\hat v(\x,\eta)$ and $\hat
\mR(\x,\eta)$? Furthermore, one may wonder how to relate the physical observables, such
as the
temperature anisotropies of the CMB, with the quantum objects that emerge from the
quantum theory? The traditional answer relies on the quantum correlation functions, in
particular, the two-point quantum correlation function $\bra 0 | \hat \mR(\x,\eta) \hat
\mR(\x',\eta) |0 \ket $ and its relation with the two-point angular correlation
function of the temperature anisotropies $\overline{\delta T/T_0 (\hat n_1) \delta T /T_0
(\hat n_2) }$, where the bar denotes an average over different directions in the
celestial sky and $\hat n_1$ and $\hat n_2$ are two unitary vectors denoting some particular
directions. We do not find the previous answer to be completely satisfactory, and for a
detailed explanation we invite the reader to consult Refs. \cite{shortcomings,
susana2013}.

In order to illustrate our approach, we begin by focusing on the
temperature anisotropies of the CMB observed today and its relation to the comoving
classical curvature  perturbation encoded in the quantity  $\mR$. Such a relation is
approximately given by  (i.e. for large angular scales)
\begin{equation}\label{deltaT}
\frac{\delta T}{T_0}  \simeq - \frac{1}{5} \mR.
\end{equation}

On the other hand, the observational data are described in terms of the
coefficients  $a_{lm}$ of the multipolar series expansion $\delta T/T_0
(\theta,\varphi)=\sum_{lm} a_{lm}Y_{lm}(\theta,\varphi)$, i.e
\beq
a_{lm}= \int
\frac{\delta T}{T_0}(\theta,\varphi)Y^*_{lm}(\theta,\varphi)d\Omega
\end{equation}
here $\theta$ and $\varphi$ are the coordinates on the celestial two-sphere,
with $Y_{lm}(\theta,\varphi)$ the spherical harmonics.

Given Eq. \eqref{deltaT}, the  coefficients $a_{lm}$ can be further re-expressed in
terms of the Fourier modes associated to $\mR$, i.e.
\begin{equation}\label{alm2_1}
  a_{lm} \simeq -\frac{4 \pi i^l}{5}   \int \frac{d^3{k}}{(2 \pi)^{3/2}} j_l (kR_D)
Y_{lm}^* (\hat{k}) \mR_{\nk},
\end{equation}
where $R_D$ is the comoving radius of the last scattering surface and $j_l (kR_D)$ the
spherical Bessel function of order $l$ of the first kind.

Finally, we can include the effects of late time physics that give rise to so called
acoustic peaks. These effects are encoded in the transfer functions $\Delta_l (k)$, and thus
the coefficients $a_{lm}$ are given by
\begin{equation}\label{alm2}
  a_{lm} = -\frac{4 \pi i^l}{5}   \int \frac{d^3{k}}{(2 \pi)^{3/2}} \Delta_l (k)
Y_{lm}^* (\hat{k}) \mR_{\nk},
\end{equation}
where $\mR_{\nk}$ is the primordial comoving curvature perturbation. Also note that for
large angular scales $\Delta_l (k) \to j_l (kR_D)$.

The next step is to relate $\mR_{\nk}$ with the quantum operator $\hat \mR_{\nk}$.
Clearly, if one computes the vacuum  expectation value $\langle 0| \hat{\mR}_{\nk} |0
\rangle$ and makes it exactly equal to  $\mR_{\nk}$, then one
obtains precisely zero; while it is clear that for
any given $l, m$, the measured value of the quantity $a_{lm}$ is not zero. As matter of
fact, the standard argument is that it is not the quantity $a_{lm}$ that is zero but the
average $\overline{a_{lm}}$. However, the notion of average is subtle, since in the CMB
one has an average over different directions in the sky, while the average that one
normally associates to the quantum expectation value of an operator is related to an
average over possible outcomes of repeatedly measurements of an observable associated to
an operator in the Hilbert space of the system (it is evident that concepts such as
measurements, observers, etc. are not well defined in the early Universe).

On the other hand, we will assume that the quantity $\mR_{\nk}$, i.e. the classical value
associated to the Fourier mode of the comoving curvature perturbation $\mR(\x,\eta)$,
is an adequate description if the quantum state associated to each mode is sharply peaked
around some particular value. In consequence, the classical value corresponds to the
expectation value of $\hat \mR$ in that particular ``peaked'' state \cite{gabriel}. In
other words, our assumption is that the
CSL  mechanism will lead to a final state such that the relation
\beq\label{igualdadchingona}
\mR_{\nk} =  \bra \Psi| \hat \mR_{\nk} |  \Psi \ket = \frac{1}{z^2} \bra \Psi | \hat
v_{\nk} | \Psi \ket
\eeq
holds.

Therefore, in our approach, the coefficients $a_{lm}$  in Eq. \eqref{alm2}, will be
given by
\begin{equation}\label{alm3}
  a_{lm} = -\frac{4 \pi i^l}{5}   \int \frac{d^3{k}}{(2 \pi)^{3/2}}
Y_{lm}^* (\hat{k}) \Delta_l (k) \bra \Psi | \hat{\mR}_{\nk} | \Psi \ket,
\end{equation}
where $| \Psi \ket$ corresponds to the evolved state according to the
non-unitary modification of the Schr\"{o}dinger equation provided by the CSL
mechanism (see Refs. \cite{jmartin,hinduesS,hinduesT} for other ways
to relate $\mR_{\nk}$ and $\hat \mR_{\nk}$ using the CSL model, and \cite{LB15} for a
discussion on those approaches). Note also that $| \Psi \ket$  does not share the
same symmetries as the vacuum state, i.e. the inhomogeneity and isotropy of the system is
encoded in the quantum state $| \Psi \ket$. Furthermore, \eq{alm3} shows how
the expectation value of the quantum field
$\hat{\mR}_{\nk}$ in the state $| \Psi \ket$ acts as a source for the coefficients
$a_{lm}$.

A well known observational quantity is the angular power spectrum defined
by
\beq
C_{l}\equiv \frac{1}{2l+1} \sum_m |a_{lm}|^2.
\eeq
We will assume that we can identify the
observed value $|a_{lm}|^2$ with the most likely value of $|a_{lm}|^2_{ML}$ obtained from
the theory and, in turn, assume that the most likely value coincides approximately with
the average $\overline{|a_{lm}|^2}$. This average is over possible realizations or
outcomes of the state $|\Psi\ket$ that results from the CSL evolution.
Thus, the observed $C_{l}^{\textrm{obs.}}$ approximately coincides with
the theoretical prediction $C_l$ given in terms of the average
$\overline{|a_{lm}|^2}$, i.e.
\beq
C_{l}^{\textrm{obs.}}  \simeq C_l =  \frac{1}{2l+1} \sum_m
\overline{|a_{lm}|^2}.
\eeq
Using Eq. \eqref{alm3}, the theoretical prediction for the angular power spectrum is
\beq\label{almavg}
C_l = \frac{1}{2l+1} \sum_m \frac{16 \pi^2}{25}   \int
 \frac{d^3{k} d^3{k'}}{(2 \pi)^3} \Delta_l (k) \Delta_l (k')  Y_{lm}^* (\hat{k}) Y_{lm}
(\hat k')  \overline{\bra \hat \mR_{\nk}  \ket \bra \hat \mR_{\nk'}  \ket^* }.
\eeq
Moreover, if the CSL evolution is such that there is no correlation between modes (which
can be justified by the fact that we are working at linear order in cosmological
perturbation theory), then
\beq
\overline{\bra \hat \mR_{\nk}  \ket \bra \hat \mR_{\nk'}  \ket^* } = \left(
\overline{\bra
\hat \mR_{\nk}^R \ket^2} + \overline{ \bra \hat
\mR_{\nk}^I \ket^2 } \right) \delta(\nk-\nk'),
\eeq
where $\hat \mR_{\nk}^{R,I} $ denotes the real and imaginary part of the field
$\hat \mR_{\nk}$ (also, we assume that there is no correlation between $\hat \mR_{\nk}^{R} $
and $\hat \mR_{\nk}^{I} $ ). Therefore,
\beq
C_l = \frac{1}{2l+1} \sum_m \frac{16 \pi^2}{25}   \int
\frac{d^3{k}}{(2 \pi)^3} |Y_{lm} (\hat k)|^2 \Delta_l^2(k)  \left( \overline{\bra
\hat \mR_{\nk}^R \ket^2} + \overline{ \bra \hat
\mR_{\nk}^I \ket^2 } \right).
\eeq

Performing the integral over the angular part of $\nk$ and summing over $m$, we obtain
\beq\label{ClCSL}
C_l = \frac{2}{25 \pi} \int dk \: k^2 \Delta_l^2 (k)  \left( \overline{\bra
\hat \mR_{\nk}^R \ket^2} + \overline{ \bra \hat
\mR_{\nk}^I \ket^2 } \right).
\eeq
On the other hand, the standard relation between the primordial power spectrum and the
$C_l$ is given by
\beq\label{Clstd}
C_l = \frac{4 \pi}{25} \int \frac{dk}{k} \:   \Delta_l^2 (k) \mP_s(k),
\eeq
where $\mP_s (k)$ is the dimensionless scalar power spectrum
defined as
\beq\label{PSdef}
\overline{\mR_{\nk} \mR_{\nk'}} \equiv \frac{2\pi^2}{k^3} \mP_s (k)
\delta(\nk-\nk')
\eeq

Henceforth, Eqs. \eqref{ClCSL} and \eqref{Clstd} imply that the power spectrum in our
approach is given by
\beq\label{PSR}
\mP_s(k) = \frac{k^3}{2 \pi^2}  \left( \overline{\bra
\hat \mR_{\nk}^R \ket^2} + \overline{ \bra \hat
\mR_{\nk}^I \ket^2 } \right).
\eeq

Note that the definition of the power spectrum, Eq. \eqref{PSdef}, is the canonical
definition when dealing with classical random fields, where the average is over possible
realizations of the random fields. In cosmology, the usual identification of the
two--point quantum correlation function $\bra 0 | \hat \mR_{\nk} \hat \mR_{\nk'} | 0
\ket$ with $\overline{\mR_{\nk} \mR_{\nk}}$ is subtle and concepts such as ergodicity,
decoherence and squeezing of the vacuum state are normally invoked.

Thus, in terms of the MS variable, the scalar power spectrum in our
approach is:
\beq\label{PSchingon}
\mP_s(k) = \frac{k^3}{ 2 \pi^2 z^2}  \left( \overline{\bra \Psi
|\hat v_{\nk}^R| \Psi \ket^2} + \overline{ \bra \Psi| \hat
v_{\nk}^I|\Psi \ket^2 } \right).
\eeq

Equation \eqref{PSchingon} is the key result from this section. It shows explicitly how
to relate the quantities obtained from the quantum theory with the observed temperature
anisotropies of the CMB. It also exhibits the difference between our approach and the
traditional one.

\section{The CSL model in quasi-matter contraction and
  inflation}\label{SecImplementation}

In this section, we will focus on the specific details of implementing the CSL model
to the QMCU and the inflationary Universe, and the main goal will be to obtain a prediction for
the power spectra.

We begin by noting that, in Eq. \eqref{PSchingon}, the predictions related to the
observational data are the objects $\bra \Psi
|\hat v_{\nk}^{R,I}| \Psi \ket$. Therefore, we will apply the CSL model to each mode of
the field and to its real and imaginary parts. As a consequence, we will assume that the
evolution of the state vector characterizing
each mode of the field, written in conformal time, is given by
\beq\label{cslmodos}
| \Psi_{\nk}^{R,I}, \eta \ket = \hat T \exp\left\{-\int_{\tau}^\eta d\eta'\left[i \hat
H_{\nk}^{R,I} +  \frac{1}{4\lambda_{k}} \left(\mW(\eta')-2\lambda_{k} \hat
\varTheta_{\nk}^{R,I} \right)^2 \right]  \right\}
|\Psi_{\nk}^{R,I}, \tau \ket
\eeq
with $H_{\nk}^{R,I}$ given in \eqref{ham}. Note that the Hamiltonian $\hat
H_{\nk}^{R,I} $depends on the field  $\hat v^{R,I}_{\nk}$ which is
defined in terms of the inflaton perturbations, but also it can be defined
analogously using the perturbations of the scalar field associated to the
QMCU [one has to take into account the change in $z(\eta)$]. Furthermore, we will
consider that Eq. \eqref{operadorcsl} can be extrapolated to a generic quantum
mode $\hat F_{\nk}$, that is, the real and imaginary parts of the mode $\hat F_{\nk}$
satisfy:
\beq\label{operadorcslmodos}
\frac{d}{d\eta} \overline{ \bra \hat F_{\nk}^{R,I} \ket } = -i \overline{[\hat
F_{\nk}^{R,I}, \hat H_{\nk}^{R,I}]} - \frac{\lambda_k}{2} \overline{[\hat
\varTheta_{\nk}^{R,I}, [\hat \varTheta_{\nk}^{R,I}, \hat F_{\nk}^{R,I}]]}.
\eeq

At this point, we have to make a choice regarding the collapse operator $\hat
\varTheta_{\nk}^{R,I}$. At first sight, the natural candidate is the MS variable, namely
$\hat
\varTheta_{\nk}^{R,I} = \hat v_{\nk}^{R,I}$. Nevertheless, we think that in absence
of a full relativistic CSL model, there is no \emph{a priori} choice and, thus, the
canonical conjugated momentum $\hat p_{\nk}^{R,I}$ can also be considered as the collapse
operator. In fact, in Ref. \cite{LB15}, we have shown that in the framework of the
inflationary Universe, the momentum operator can be used as the collapse operator given
that, in the longitudinal gauge, the momentum operator is directly related with
the curvature perturbation.

Thus, we are going to consider four different cases:

\begin{enumerate}
  \item[(i)] The collapse operator is $ \hat v_{\nk}^{R,I}$  during slow roll inflation.

  \item[(ii)] The collapse operator is $\hat v_{\nk}^{R,I}$  during the
QMCU.

  \item[(iii)] The collapse operator is $\hat p_{\nk}^{R,I}$ during slow roll inflation.

  \item[(iv)] The collapse operator is $\hat p_{\nk}^{R,I}$  during the
QMCU.

\end{enumerate}

We stress that only the third case, that is, the implementation of the CSL
model within the inflationary framework using the field $ \hat {p}_{\nk}^{R,I}$  as the
collapse operator, was first developed in Ref. \cite{LB15}. Nevertheless, we are
including it in the present work for the sake of completeness. Note however
that the analysis in  Ref. \cite{LB15}  was done in the longitudinal gauge. In the
present paper, we will work in the comoving gauge in all the four cases. The analysis of the
three remaining cases, and in particular the implementation of the CSL model during a
contracting phase of the early Universe, are presented here for the first time.

For each of these four cases, we will obtain the scalar (and tensor) power spectrum.

Furthermore, the calculation of the object $\overline{\bra \Psi
  |\hat v_{\nk}^{R}| \Psi \ket^2}$ is identical to $\overline{\bra \Psi
  |\hat v_{\nk}^{I}| \Psi \ket^2}$. Consequently, we will omit from now on the indexes
$R,I$ unless it creates confusion.

Using the Gaussian wave functions in the field representation, Eq.
\eqref{psiondav}, and the probability associated to $\mW(\eta)$ in Eq. \eqref{prob}, it
can be shown that \cite{pedro},
\beq\label{restav}
\overline{\bra \hat v_{\nk} \ket^2} = \overline{\bra \hat v_{\nk}^2 \ket} -
\frac{1}{4 \text{Re}[A_k (\eta)]}.
\eeq

The quantity $(4 \text{Re}[A(\eta)])^{-1}$ is the standard deviation of the squared
field variable $\hat v_{\nk}$. It is also the width of every packet in Fourier's space.
In a similar manner, using the Gaussian wave function in the momentum representation,
Eq. \eqref{psiondap}, along with Eq. \eqref{prob}, it follows that
\beq\label{restap}
\overline{\bra \hat v_{\nk} \ket^2} = \overline{\bra \hat v_{\nk}^2 \ket} -
\frac{|\tilde A_k (\eta)|^2}{ \text{Re}[\tilde A_k (\eta)]}.
\eeq

For cases (i) and (ii), it is convenient to work with Eq. \eqref{restav}; and for
cases (iii) and (iv) with Eq. \eqref{restap}. Thus, to
calculate $\overline{\bra \hat v_{\nk} \ket^2}$, we only need to find the two
terms on the right hand side of \eqref{restav} or
\eqref{restap}, respectively. The second term on the right hand side of both equations can be
found from the CSL evolution equation, Eq. \eqref{cslmodos}, while the first one by using
Eq. \eqref{operadorcslmodos} with the wave function in the corresponding
representation. Also, in \eqs{restav} and \eqref{restap}, we consider the regime
$-k\eta \to 0$, which correspond to the range of observational interest, that is, the
regime for which the modes are larger than the Hubble radius.

Once we have computed Eqs. \eqref{restav} and \eqref{restap}, in the corresponding case,
we can substitute it into Eq. \eqref{PSchingon} to give a specific prediction for the
scalar power spectrum. The actual calculations are long, so we have included them in
Appendix \ref{app} for the interested reader. In the following, we will show only the main results.

In case (i), our predicted scalar power spectrum during inflation is
(at the lowest order in the slow roll parameter):
\beq\label{PSescalarcaso1}
P_s(k) = \frac{H^2 (-k\eta)^{-2\nu_s+3}}{\pi \epsilon M_P^2} F_1(\lambda_k,\nu_s),
\eeq
with $\nu_s \equiv {3}/{2} + 2 \epsilon - \delta$ and
\barr\label{F1}
F_1(\lambda_k,\nu_s) &\equiv&
\frac{2^{2\nu_s-3}}{\sin^2(\nu_s \pi) \Gamma^2(1-\nu_s)}   \bigg[ 1- \frac{\lambda_k
  \tau}{k} + \frac{3 \lambda_k}{k^2} \sin (-k\tau) \cos (-k\tau)  \bigg]    \nn
&-& \frac{1}{8\pi} \left[ \frac{\lambda_k}{2(\nu_s-1)k^2} (-k\eta)^{-2\nu_s+2} +
\frac{\zeta_k^{2\nu_s}\pi \sin(\pi \nu_s + 2 \nu_s \theta_k)    }{\sin (\pi \nu_s)
2^{2\nu_s}\Gamma^2 (\nu_s) }        \right]^{-1}.
\earr

For case (ii), we have
\beq\label{PSescalarcaso2}
P_s(k) = \frac{1}{12 \pi^2}  \left( \int_{-\infty}^\eta
\frac{d\tilde \eta}{z^2} \right)^2 \left(\frac{k}{|aH|} \right)^{-2\mu_s+3} F_2
(\lambda_k,\mu_s),
\eeq
with $\mu_s \equiv {3}/{2}- 6 \bar \epsilon $ and
\barr\label{F2}
 F_2(\lambda_k,\mu_s) &\equiv& \frac{8 \pi}{\sin^2(\mu_s \pi) \Gamma^2(1-\mu_s)}   \bigg[
1-
\frac{\lambda_k
  \tau}{k} - \frac{3 \lambda_k}{k^2} \sin (-k\tau) \cos (-k\tau)  \bigg]    \nn
&-&  \left[ \frac{2^{2\mu_s-4}\lambda_k}{(\mu_s-1)k^2} (-k\eta)^{-2\mu_s+2} +
\frac{\zeta_k^{2\mu_s}\pi \sin(\pi \mu_s + 2 \mu_s \theta_k)    }{\sin (\pi \mu_s)
8\Gamma^2 (\mu_s) }        \right]^{-1}.
\earr

In both cases, (i) and (ii), we have also defined:
\beq\label{zetakv}
\zeta_k \equiv \left(1+\frac{4\lambda_k^2}{k^4} \right)^{1/4}, \qquad \qquad \theta_k
\equiv
-\frac{1}{2} \arctan \left( \frac{2\lambda_k}{k^2} \right).
\eeq

The calculations for obtaining the tensor power spectra are very similar to the one used
to obtain the scalar ones (see Appendix \ref{app} for further details). In case (i), the
formula obtained for the tensor power spectrum is
\beq\label{PStensorcaso1}
P_t(k) = \frac{H^2 16 (-k\eta)^{-2\nu_t+3}}{\pi M_P^2} F_1 (\lambda_k,\nu_t),
\eeq
where $\nu_t \equiv {3}/{2} + \epsilon$. Therefore, the tensor-to-scalar ratio $r
\equiv P_t(k)/P_s(k)$, at the lowest order in the slow roll parameter, is given by
\beq\label{rinflacion}
r = 16 \epsilon
\eeq
which is exactly the same prediction as in the standard inflationary slow roll scenario.

Meanwhile, in case (ii), the tensor power spectrum is
\beq\label{PStensorcaso2}
P_t(k) = \frac{2}{9 \pi^2}  \left( \int_{-\infty}^\eta
\frac{d\tilde \eta}{z_T^2} \right)^2 \left(\frac{k}{|aH|} \right)^{-2\mu_t+3} F_2
(\lambda_k,\mu_t),
\eeq
where $ \mu_t \equiv {3}/{2} - 6 \bepsilon = \mu_s$. Also, for very low
energy densities and curvatures, $z_T=a$ (see Ref. \cite{elizalde}).
The  tensor-to-scalar ratio is given by
\beq\label{rQMCU}
r = \frac{8}{3} \left( \frac{\int_{-\infty}^\infty
  \frac{d\eta}{z_T^2(\eta)}}{\int_{-\infty}^\infty \frac{d\eta}{z^2(\eta)}}
\right)^2_{k=|aH|},
\eeq
which is also the same as the one presented in Refs. \cite{elizalde,jaume}. Note that we
have evaluated the upper limit of the integrals at $\eta =\infty$. The motivation  is
essentially the same as the one given in Refs. \cite{elizalde,jaume}. That is, one
evaluates the scalar and power spectra at very late times corresponding to when the mode
``re-enters the horizon'', or more precisely when $k = |aH|$ during the expanding
(post-bounce) phase.

The previously presented cases (i) and (ii) correspond to selecting $\hat v_{\nk}^{R,I}$
as the collapse operator. Next, we focus on the results for cases (iii) and (iv), which
correspond to choose $\hat p_{\nk}^{R,I}$ as the collapse operator.

For case (iii), we obtain:
\beq\label{PSescalarcaso3}
P_s(k) = \frac{2^{2\nu_s-3}H^2 \Gamma^2(\nu_s)}{ \epsilon M_P^2 \pi^3}
(-k\eta)^{-2\nu_s+1} F_3 (\lambda_k,\nu_s),
\eeq
where we have defined $\nu_s \equiv {1}/{2}+2\epsilon-\delta$ and
\beq\label{F3}
F_3 (\lambda_k,\nu_s) \equiv 1- \lambda_ k k \tau + \lambda_k \sin (-k\tau) \cos (-k\tau)
- \frac{\sin(\pi \nu_s)}{\tilde \zeta^{2\nu_s} \sin(2\nu_s \tilde \theta_k + \pi \nu_s)}.
\eeq

In case (iv), the corresponding expression results
\beq\label{PSescalarcaso4}
P_s(k) = \frac{1}{12 \pi^2}  \left( \int_{-\infty}^\eta
\frac{d\tilde \eta}{z^2} \right)^2 \left(\frac{k}{|aH|} \right)^{-2\mu_s+3}
F_4(\lambda_k,\mu_s)
\eeq
with the definitions $\mu_s \equiv 3/2-6\bepsilon$ and
\barr\label{F4}
F_4(\lambda_k,\mu_s) &\equiv& \frac{8\Gamma^2(\mu_s)}{\pi}   \bigg[ 1- \lambda_ k k
\tau -5 \lambda_k
\sin (-k\tau) \cos (-k\tau) \bigg]    \nn
&+& \frac{c_1 2^{-2\mu_s+4}(1+4\lambda_k)^2}{2\lambda_k (-k\eta)^{-2\mu_s} +c_2
  4\lambda_k (-k\eta)^{-2\mu_s+2} +c_3 \tilde \zeta_k^{2\mu_s}
\sin[2(\mu_s+1)\tilde \theta_k + \pi
  \mu_s]}.
\earr
The constants $c_1,c_2$ and $c_3$ are shown in Appendix \ref{app}. In
both cases, (iii) and (iv), we have the following definitions
\beq\label{zetakp}
\tilde{\zeta}_k \equiv (1+4\lambda_k^2)^{1/4}, \qquad \tilde{\theta_k} \equiv -
\frac{1}{2} \arctan (2\lambda_k).
\eeq

The predictions for the tensor-to-scalar ratios are exactly the same as the ones presented in
cases (i) and (ii) (see Appendix \ref{app}).

We end this section by summarizing the main results. We have applied the CSL model to the
inflationary Universe and to the QMCU. Moreover, in order to employ the CSL model, we
need to choose the collapse operator. We have chosen to work with $\hat v_{\nk}^{R,I}$
and $\hat p_{\nk}^{R,I}$ as the collapse operators. Henceforth, we have obtained the
scalar power spectra in  four different cases \eqs{PSescalarcaso1},
\eqref{PSescalarcaso2}, \eqref{PSescalarcaso3} and \eqref{PSescalarcaso4}. On the other
hand, introducing the CSL mechanism does not affect the tensor-to-scalar ratio $r$.
Specifically, if one works within the standard inflationary scenario, then the prediction
for $r$ is equal to the standard one given by slow roll inflation; meanwhile, if one
adopts the QMCU framework, then the predictions are equal to the ones presented in Refs.
\cite{elizalde,jaume}.

\section{Discussion on the CSL inspired power spectra}
\label{discussionPS}

In this section, we will discuss the implications of the results obtained in the previous
section. In particular, we will compare our predicted scalar power spectra with the
standard one.

The scalar power spectrum predicted by slow roll inflation is traditionally expressed as
\cite{mukhanov1992,mukhanovbook}
\beq\label{PScanonico}
P_s(k) = A_s \left( \frac{k}{k_0} \right)^{n_s-1},
\eeq
where $k_0$ is a pivot scale, and the amplitude $A_s$ and the spectral index $n_s$ are given by
\beq\label{defAsns}
A_s = \left( \frac{H^2}{8 \pi^2 M_P^2 \epsilon}  \right)_{k_0=aH}, \qquad n_s-1 =
-4\epsilon + 2\delta.
\eeq

On the other hand, we have four different expressions for the scalar power spectrum,
corresponding to the four cases mentioned at the beginning of Sect.
\ref{SecImplementation}. In the following, we will analyze each one of them, but first we
will make a few observations regarding the parameter $\lambda_k$.

The dependence on $k$ in the parameter $\lambda_k$ encodes the ``amplification
mechanism'', which is characteristic of dynamical reduction models (see Refs.
\cite{hinduesS, hinduesT}). One possible way to determine the exact
dependence on $k$, and perhaps the simplest, is by dimensional analysis. That is, the
main evolution equations are given in \eqs{cslmodos} and \eqref{operadorcslmodos}; consequently, in order for those
equations to be dimensionally consistent,  the fundamental dimensions of $\lambda_k$
change depending on the fundamental units associated to the collapse operator $\hat
\varTheta_{\nk}^{R,I}$.  Moreover, we expect that $\lambda_k$ is directly related to
$\lambda$, i.e. the CSL parameter, which clearly must be the same in all physical
situations (cosmological or otherwise). Moreover, taking into account that  we are
working in units in which $\hbar = c=1$, the fundamental dimension of $\lambda$ is [Length]$^{-1}$.

Thus, in the case where the collapse operator is chosen to be
$\hat v_{\nk}^{R,I}$, the most natural expression of $\lambda_k$, which is consistent
with the dimensions of all terms involving  the dynamical equations, is
\beq\label{vlambdak}
\lambda_k = \lambda k
\eeq
And in the case where the selected collapse operator is $\hat p_{\nk}^{R,I}$, such an expression is
\beq\label{plambdak}
 \lambda_k = \frac{ \lambda}{k},
\eeq
where $\lambda$ is the CSL parameter, with the same numerical value in all
cases. From now on, we will assume that $\lambda_k$ takes the form of \eqs{vlambdak} and
\eqref{plambdak} depending on the chosen operator acting as the collapse operator.

\subsection{The CSL power spectra during inflation}

Let us begin the discussion by working within the framework of the inflationary Universe, analyzing cases (i) and (iii).

The scalar power spectrum given in \eq{PSescalarcaso1}, corresponding to case (i), can be written in a similar form to the one showed in \eq{PScanonico}. As usual, the power spectrum can be evaluated at the conformal time where the pivot scale
``crosses the horizon''; or more precisely, when $-k_0 \eta = 1$ (i.e. $k_0=
aH$) during the inflationary epoch. Furthermore, the different coefficients that multiply
each term of the function $F_1(\lambda_k,\nu_s)$ involve the quantity $\nu_s$. For these
terms, we can approximate $\nu_s \simeq 3/2$ without loss of generality. However, note
that such approximation cannot be done to the powers of $k$ involving $\nu_s$ because these
are directly related to the scalar spectral index $n_s$, for which the value $n_s=1$ is
ruled out. Furthermore, in order to provide a suitable normalization for the CSL power
spectra, we multiply and divide by the quantity $\lambda| \tau|$.
Thus, the power spectrum in \eq{PSescalarcaso1} can be rewritten as:
\beq\label{PScaso1}
P_s(k) = A_s \left( \frac{k}{k_0} \right)^{n_s-1} C_1(k)
\eeq
where
\beq\label{Asynsinflation}
A_s = \left( \frac{H^2 \lambda |\tau| }{4 \pi^2 M_P^2 \epsilon} \right)_{k_0 = aH} ,
\qquad
n_s-1 =
-4\epsilon + 2\delta,
\eeq
and $C_1(k) \equiv F_1 (\lambda_k = \lambda k, \nu_s \simeq 3/2) /\lambda |\tau|$; that
is,
\beq\label{C1}
C_1(k) = \frac{1}{\lambda |\tau|}  \left\{ 1- {\lambda
  \tau} + \frac{3 \lambda}{k} \sin (-k\tau) \cos (-k\tau)   -\frac{1}{2} \left[
\frac{\lambda}{k} (-k\eta)^{n_s-2} +
\frac{\zeta_k^{4 - n_s} \cos[ (4 - n_s ) \theta_k]    }{2}        \right]^{-1} \right\}
\eeq
[expressions for $\zeta_k$ and $\theta_k$ are given in \eq{zetakv} with $\lambda_k =
\lambda k$].

Within the inflationary framework, and with the same arguments followed to arrive to \eq{PScaso1},
we can write the power spectrum \eq{PSescalarcaso3}, corresponding to case (iii), in the following form:
\beq\label{PScaso3}
P_s(k) = A_s \left( \frac{k}{k_0} \right)^{n_s-1} C_3(k)
\eeq
where $A_s$ and $n_s$ are the same as in \eq{Asynsinflation}, and $C_3(k) \equiv F_3
(\lambda_k = \lambda/k, \nu_s \simeq 1/2) /\lambda |\tau|$. Thus,
\beq\label{C3}
C_3(k) =  \frac{1}{\lambda |\tau|} \left\{ 1 - \lambda \tau + \frac{\lambda}{k}
\sin(-k\tau) \cos(-k\tau) -
\frac{1}{\tilde   {\zeta}_k^{2 -n_s} \cos[(2 -n_s)\tilde\theta_k]   } \right\}
\eeq
[expressions for $\tilde \zeta_k$ and $\tilde \theta_k$ are given in \eq{zetakp} with
$\lambda_k = \lambda/k$]

Let us make some remarks. Notice that the scalar index predicted by the CSL power
spectra is exactly the same as the standard one from slow roll inflation, but the amplitude is slightly different.
The difference between the standard amplitude and the one using the CSL model is a factor of $\lambda |\tau|/2$ [see
Eqs. \eqref{defAsns} and \eqref{Asynsinflation}]. The reason for the factor $1/2$ can be traced back
to \eq{PSchingon}, since in our approach the power spectrum receives an equal contribution from the
expectation values $\overline{ \bra \hat v_{\nk}^R \ket^2 }$ and $\overline{ \bra \hat
v_{\nk}^I \ket^2 }$. However, the factor $1/2$ will not have any important observational consequences. On the other hand, the factor
$\lambda |\tau|$, which comes from the normalization of $C_1(k)$ and $C_3(k)$, does modify
the standard predicted amplitude. A quantitative analysis will be done in the next
section.

A second remark has to do with the following. It is well known that there is a minimum number of $e$-foldings for inflation related to the
solution of the ``horizon problem'', and this minimum number depends on the characteristic energy of
inflation. A shared characteristic of the functions $C_1(k)$ and $C_3(k)$ is that they include the
quantity $\tau$, which represents the conformal time at the beginning of inflation. This
quantity depends on the energy scale at which inflation ends, which is associated to the
inflaton potential $V$ at that time, and the number of $e$-foldings corresponding to the total duration of inflation.

Third, note that another important feature of the CSL power spectra in inflation is that the function
$C_1(k)$, corresponding to the case in which the collapse operator is $\hat
v_{\nk}^{R,I}$, depends explicitly on the conformal time $\eta$, whilst the function
$C_3(k)$, which corresponds to the case when $\hat p_{\nk}^{R,I}$ is the collapse
operator, does not exhibit such time dependence. The time dependence on the power
spectrum when the collapse operator is $\hat v_{\nk}^{R,I}$ has been previously noted by other
authors \cite{jmartin,hinduesS}. Nevertheless, the exact form of their predicted power
spectrum is different from the one shown here. As a matter of fact, that difference
is illustrated by considering the limiting case $\lambda_k=0$. In such mentioned
works, for $\lambda=0$ (i.e. standard Schr\"odigner evolution), their predicted power
spectrum is the same as the traditional one. Contrarily, in our approach, if $\lambda_k=0$ then
$P_s(k)=0$, which is consistent with our point of view regarding the role played by the
CSL model. In any case, even if the pictures used for the role of the CSL model are
different between our work and the one in Refs.  \cite{jmartin,hinduesS}, the time dependence on the
power spectrum is shared.

In order to continue, we choose to evaluate the power spectrum (or
equivalently the function $C_1(k)$) at the conformal time when inflation ends, which we
denote by $\eta_f$. We think it is consistent with the previous calculations in
which the power spectrum was obtained in the limit $-k\eta \to 0$, which is satisfied by
the value $\eta_f$. The precise value of $\eta_f$ depends mainly on the
characteristic energy scale of inflation and the number of $e$-foldings assumed for the
full inflationary phase $N \equiv\ln [a(\eta_f)/a(\tau)]$.

Readers familiar with previous works, can check that our expression for the scalar power spectrum \eq{PScaso3}, which features the
function $C_3(k)$, is essentially the same as the one obtained in Ref. \cite{LB15}. The
difference is that in the present paper we chose to work in the comoving gauge (where
$\mR$ represents the curvature perturbation), whilst in the aforementioned
reference we worked in the longitudinal gauge, where the Bardeen potential $\Phi$
corresponds to the curvature perturbation. Therefore, we find reassuring that even having worked
in different gauges, the expression for the power spectrum, when the collapse
operator is the momentum associated to the Fourier's mode of the MS variable, is the
same and it has the attractive feature that does not depends on the conformal time.

\begin{figure}
\begin{center}
\includegraphics[scale=1.0]{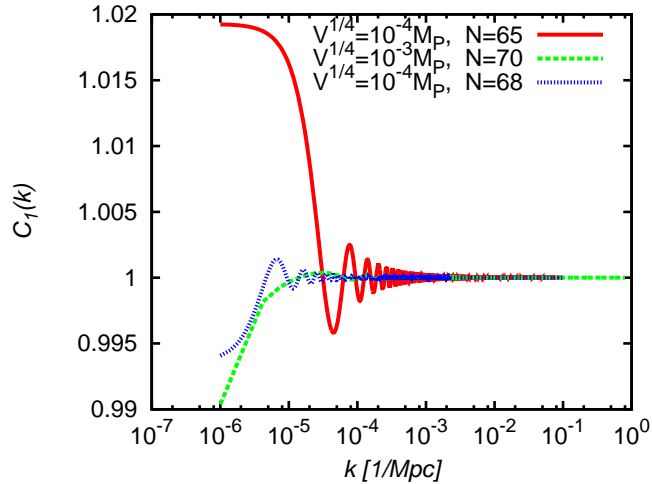}
\end{center}
\caption{The function $C_1(k)$, corresponding to the power spectrum in the case when the
collapse operator is $\hat v_{\nk}^{R,I}$ during inflation. We have set the value
$\lambda_{\textrm{GRW}} = 1.029 \times 10^{-2}$ Mpc$^{-1}$ and $n_s=0.96$. The values of conformal
time at the beginning of the inflationary regime $\tau$ corresponding to the three cases depicted (from top to bottom)
are: $-\tau =7803894, 115820063$ and $156745414$ Mpc. We have
evaluated $C_1(k)$ at $\eta = \eta_f$, and the values of $\eta_f$
corresponding to the three different cases considered (from top to bottom) are:
$-\eta_f = 4.604 \times 10^{-22}, 4.604 \times 10^{-23}$ and $4.604 \times 10^{-22}$ Mpc.
$M_P$ denotes the reduced Planck mass $M_P \simeq 10^{18}$ GeV. }
\label{plotCk1}
\end{figure}

\begin{figure}
\begin{center}
\includegraphics[scale=1.0]{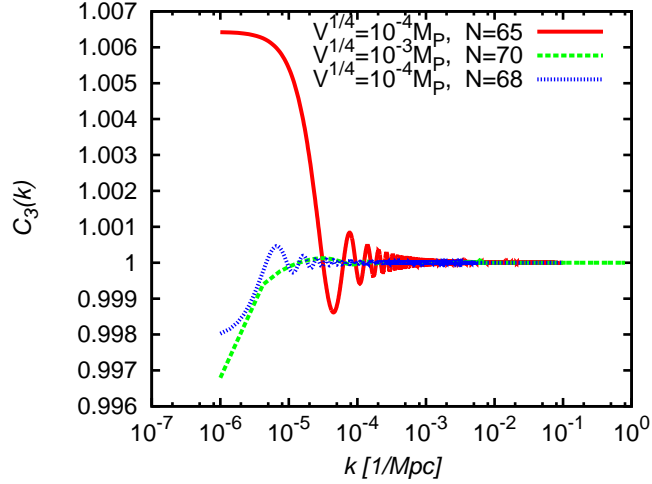}
\end{center}
\caption{The function $C_3(k)$, corresponding to the power spectrum in the case when the
collapse operator is $\hat p_{\nk}^{R,I}$ during inflation. We have set the value
$\lambda_{\textrm{GRW}} = 1.029 \times 10^{-2}$ Mpc$^{-1}$ and $n_s=0.96$. The values of conformal
time at the beginning of the inflationary regime $\tau$ corresponding to the three cases depicted (from top to bottom)
are: $-\tau =7803894, 115820063$ and $156745414$ Mpc. $M_P$ denotes
the reduced Planck mass $M_P \simeq 10^{18}$ GeV.}
\label{plotCk3}
\end{figure}

Figures \ref{plotCk1} and \ref{plotCk3} show different plots for the functions
$C_1(k)$ and $C_3(k)$, respectively. In both cases, we have considered the value of the CSL parameter as
$\lambda_{\textrm{GRW}} = 1.029 \times 10^{-2}$ Mpc$^{-1}$, which corresponds to a value favored by experimental
data \cite{mohammed,mohammed2,bassibounds2016,bassibounds22016}. The various
plots in each figure correspond to different values of the
characteristic energy of inflation $V^{1/4}$, and the total $e$-foldings $N$ that
inflation is assumed to last, which also set the values of $\tau$ and $\eta_f$. The values of $k$
considered correspond to these of observational interest, i.e. we consider $k$ in
the range from $10^{-6}$ to $10^{-1}$ Mpc$^{-1}$.

As we can observe, the functions $C_1(k)$ and $C_3(k)$
exhibit an oscillatory behavior around the unit. For increasing values of $k$, the
oscillations decrease in amplitude. However, we note that even for decreasing values of
$k$ the functions $C_1(k)$ and $C_3(k)$  are very close to 1. Consequently, for the
chosen values of $\lambda$, $V^{1/4}$ and $N$, the functions $C_1(k) \simeq C_3(k)
\simeq1$. That means that the shape of the angular power spectrum $C_l$ will not
be very different from the standard one, but the amplitude could vary (a complete analysis
will be presented in the next section).

Additionally, the fact that $C_1(k)$ depends on the conformal time does not seem to
affect its behavior in a significant manner. In fact, it is closely similar to the one of
$C_3(k)$, which does not depend on the conformal time. That means that the contribution
from the time dependent term (i.e. the last term in \eq{C1}), to the
total value of the function $C_1(k)$ is negligible when $-k\eta \to 0$.

\subsection{The CSL power spectra in the QMCU}

We switch now the discussion to the framework of the QMCU, i.e. cases (ii) and
(iv), which correspond to selecting $\hat v_{\nk}^{R,I}$ or $\hat
p_{\nk}^{R,I}$ as the collapse operator, respectively.

The scalar power spectra given in both cases, i.e. \eqs{PSescalarcaso2} and
\eqref{PSescalarcaso4}, can also be written in a manner similar to the standard spectrum
\eq{PScanonico}. Once again, following Refs. \cite{jaume,elizalde}, we choose to evaluate
the spectrum at the conformal time
where the pivot scale ``reenters the horizon'' $k_0 = |aH|$, which happens at late times
during the expansion phase of the Universe (consequently the upper limit of the integral
is evaluated at $\eta\to \infty$). We approximate (for the same arguments as in the
previous subsection) $\mu_s \simeq 3/2$ in the coefficients of the terms in expressions
$F_2(\lambda_k,\mu_s)$ and $F_4(\lambda_k,\mu_s)$ (but not in the powers of $k$ as
these powers are directly related to the spectral index $n_s$). Additionally, the
parameter $\lambda_k$ is assumed to be $\lambda_k = \lambda k$ for case (ii) and
$\lambda_k = \lambda/k$ in case (iv). Moreover, we multiply and divide by a factor
of $\lambda |\tau|$ in order to properly normalize the expressions $F_2$ and $F_4$.

Henceforth, the scalar power spectrum for case (ii), \eq{PSescalarcaso2}, will be
written as
\beq\label{PScaso2}
P(k) = A_s \left( \frac{k}{k_0} \right)^{n_s-1} C_2(k)
\eeq
where
\beq\label{Aynscontraccion}
A_s = \frac{\lambda |\tau|}{6 \pi^2} \left( \int_{-\infty}^{\infty}
\frac{d\eta}{z(\eta)^2}   \right)^2,    \qquad n_s-1 = 12 \bepsilon
\eeq
and $C_2(k) \equiv F_2(\lambda_k = \lambda k,\mu_s \simeq 3/2)/\lambda |\tau|$; thus,
\beq\label{C2}
C_2(k) = \frac{1}{\lambda |\tau|} \left\{         \bigg[
1- \lambda \tau - \frac{3 \lambda}{k} \sin (-k\tau) \cos (-k\tau)  \bigg]   - \frac{1}{2}
\left[ \frac{\lambda}{k} (-k\eta)^{n_s-2} +
\frac{\zeta_k^{4-n_s}}{2} \cos[(4-n_s) \theta_k]          \right]^{-1} \right\}.
\eeq

On the other hand, the power spectrum for case (iv), \eq{PSescalarcaso4}, will be written as
\beq\label{PScaso4}
P(k) = A_s \left( \frac{k}{k_0} \right)^{n_s-1} C_4(k)
\eeq
with $A_s$ and $n_s$ the same as in \eq{Aynscontraccion}, and $C_4(k) \equiv F_4(\lambda_k
= \lambda/k,\mu_s \simeq 3/2)/\lambda |\tau|$. Hence,
\barr\label{C4}
C_4(k) &=& \frac{1}{\lambda |\tau|} \bigg\{   \bigg[ 1- \lambda \tau -\frac{5\lambda}{k}
\sin (-k\tau) \cos (-k\tau) \bigg]  \nn
&-& \frac{(1+4\lambda/k)^2}{6(\lambda/k) (-k\eta)^{n_s-4} + 4(\lambda/k) (-k\eta)^{n_s-2}
+
\tilde   \zeta_k^{4-n_s} \cos[\tilde \theta_k(6-n_s)    ]        }\bigg\}.
\earr

\begin{figure}
\begin{center}
\includegraphics[scale=1.0]{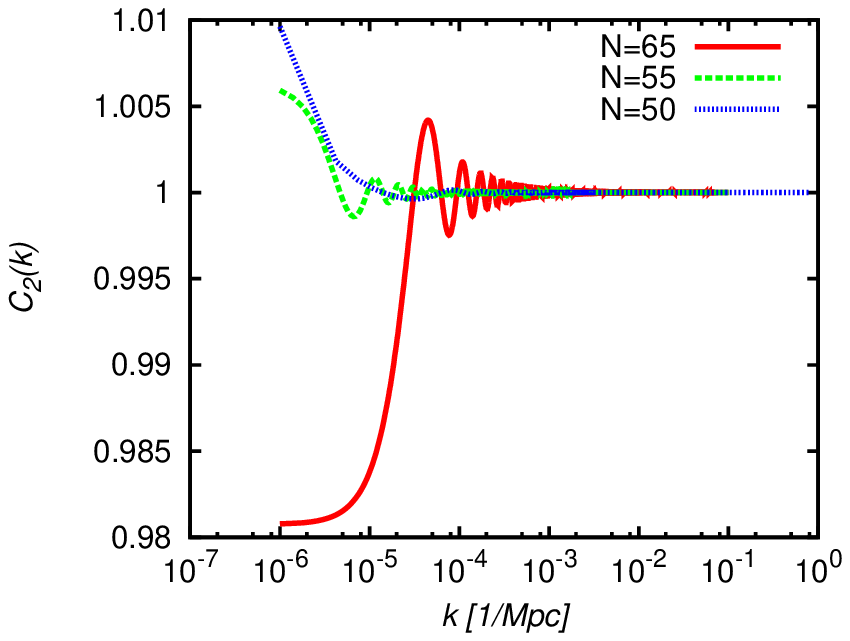}
\end{center}
\caption{The function $C_2(k)$, corresponding to the power spectrum in the case when the
collapse operator is $\hat v_{\nk}^{R,I}$ in the QMCU framework. We have set the value
$\lambda = 1.028 \times 10^{-2}$ Mpc$^{-1}$ and $n_s=0.96$. The value of conformal time at the
beginning of the quasi-matter dominated stage $\tau$ corresponding to the three cases
depicted (from top to bottom) are: $-\tau =7803894, 156745414$ and $115820063$ Mpc. We have
evaluated $C_2(k)$ at $\eta = \eta_f$, and the values of $\eta_f$ corresponding to the three
different cases considered (from top to bottom) are: $-\eta_f = 5.99 \times 10^{-8},
1.78 \times 10^{-4}$ and $1.60 \times 10^{-3}$ Mpc.}
\label{plotCk2}
\end{figure}

\begin{figure}
\begin{center}
\includegraphics[scale=1.0]{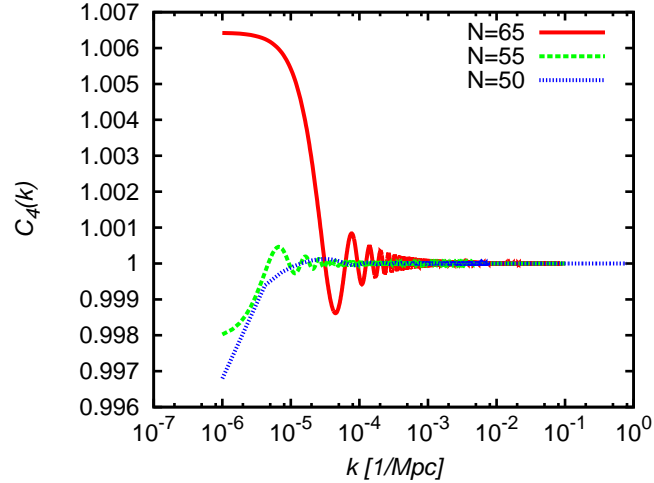}
\end{center}
\caption{The function $C_4(k)$, corresponding to the power spectrum in the case when the
collapse operator is $\hat p_{\nk}^{R,I}$ in the QMCU framework. We have set the value
$\lambda = 1.028 \times 10^{-2}$ Mpc$^{-1}$ and $n_s=0.96$. The value of conformal time at the
beginning of the quasi-matter dominated stage $\tau$ corresponding to the three cases
depicted (from top to bottom) are: $-\tau =7803894, 156745414$ and $115820063$ Mpc. We have
evaluated $C_4(k)$ at $\eta = \eta_f$, and the values of $\eta_f$ corresponding to the three
different cases considered (from top to bottom) are: $-\eta_f = 5.99 \times 10^{-8},
1.78 \times 10^{-4}$ and $1.60 \times 10^{-3} $ Mpc. }
\label{plotCk4}
\end{figure}

As in the case of the inflationary Universe, the predicted value of the scalar spectral
index $n_s$ is not affected by the CSL model. In fact, it has the same expression
as that of the QMCU original models presented in Refs. \cite{jaume,elizalde}.
Nevertheless, as can be seen in \eq{Aynscontraccion}, the amplitude of the spectrum is
modified by an extra factor of
$\lambda |\tau|/2$ with respect to the original QMCU model, that is,

\beq
A_s = A_s^{\textrm{orig.}} \times  \frac{\lambda|\tau|}{2}
\eeq

In this case, $\tau$
corresponds to the beginning of the quasi-matter dominated period. Regarding the
amplitude of the spectrum in the QMCU model, when the background
evolution is driven by a matter dominated Universe, it can be obtained
analytically working within $F(T)$ gravity or LQC. In the teleparallel $F(T)$ case, the
original
amplitude is
\beq\label{Asoriginal1}
A_s^{\textrm{orig.}} = \frac{1}{3 \pi^2} \left( \int_{-\infty}^{\infty}
\frac{d\eta}{z(\eta)^2}   \right)^2 = \frac{\pi^2}{9} \frac{\rho_c}{\rho_P}
\eeq
while in the LQC case, the original amplitude is
\beq\label{Asoriginal2}
A_s^{\textrm{orig.}} = \frac{1}{3 \pi^2} \left( \int_{-\infty}^{\infty}
\frac{d\eta}{z(\eta)^2}   \right)^2 = \frac{16}{9} \frac{\rho_c}{\rho_P} \mathcal{C}^2
\eeq
where $\rho_P$ is the Planck energy density, $\mathcal{C}\simeq 0.9159$ is Catalan's
constant and $\rho_c$ is called the critical density, which corresponds to the energy
density at which the Universe bounces, both expressions for the amplitude can be
consulted in Refs. \cite{jaume,elizalde}.

Thus, in order
to obtain an amplitude in both cases (the teleparallel gravity case and the LQC case)
that is consistent with that obtained from the CMB data (i.e. $A_s \simeq 10^{-9}$),
and taking into account the extra factor of $\lambda |\tau|/2$ coming from the CSL model, the value of the energy
density at the bouncing point must satisfy

\beq
\rho_c \simeq  10^{-9} \frac{\rho_P}{\lambda|\tau|}.
\eeq

Generically $\lambda |\tau| \gg 1$; hence, the CSL model introduces an
extra constriction to the QMCU, that is, $\rho_c \ll \rho_P$. In the next section we will
perform a more quantitative analysis.

The functions $C_2(k)$ and $C_4(k)$ share a characteristic feature, namely they depend
explicitly on $-k\eta$, which comes from a series expansion around $-k\eta \to 0$.
Consequently, we choose to evaluate $\eta = \eta_f$ corresponding to the end of the
quasi-matter domination stage or the onset of the bouncing phase. One can also define the
total number of e-foldings $N \equiv  \ln[a(\tau)/a(\eta_f)]$ for the duration of the
quasi-matter dominated phase; however, notice that in this case, since there is no horizon
problem, there is no minimum value of $N$. Another important aspect is that if
$\lambda=0$ then $C_2(k)=C_4(k) =0$. In other words, if the evolution of the state vector
is completely unitary, then there are no perturbations of the spacetime at all and the
state vector continues being perfectly symmetric, which is consistent with our
conceptual framework.

In Figs. \ref{plotCk2} and \ref{plotCk4}, we show different plots for the functions
$C_2(k)$ and $C_4(k)$, respectively. In both cases, we have considered the value
$\lambda_{\textrm{GRW}} = 1.029 \times 10^{-2}$ Mpc$^{-1}$. The various plots in each
figure correspond to different
values of  $\tau$ and $\eta_f$. The values of $k$ considered correspond to the values of
observational interest, hence, we consider $k$ in the range from $10^{-6}$ to $10^{-1}$ Mpc$^{-1}$. As
we can see, the functions $C_2(k)$ and $C_4(k)$ exhibit the same oscillatory behavior
around the unity as its counterparts during inflation. Also, the amplitude of each
oscillation decreases for increasing values of $k$.

We end this section with a few comments regarding the dependence of the power spectrum on $\eta$ in cases (i), (ii) and (iv).
In case (i), which corresponds to selecting the MS
variable as the collapse operator during inflation, the term containing the $\eta$
dependence is the last one of $C_1(k)$ in \eq{C1}.  On the other hand, since the
amplitude associated to the modes $\mR_{\nk}$ is ``frozen'' on super-Hubble scales,
the behavior of $C_1(k)$ will not change for super-horizon modes. As a
matter of fact, the plots in Fig. \ref{plotCk1} show
that $C_1(k)$ is essentially a constant in the limit $-k\eta \to 0$, which means that
the term containing the $\eta$ dependence is sub-dominant in such a limit. In cases (ii)
and (iv), corresponding to the framework of the QMCU, the behavior of the functions
$C_2(k)$ and $C_4(k)$ are very similar to that of $C_1(k)$ (see Figs. \ref{plotCk2} and \ref{plotCk4}).
That is, they are practically a constant in the limit $-k\eta \to0$, which
means that the terms involving $\eta$, i.e. the last terms of \eqs{C2} and
\eqref{C4}, are sub-dominant in the super-Hubble limit. Nevertheless, since in cases (ii)
and (iv) the Universe approaches a non-singular bounce, it might be the case that, when
the mode ``reenters the horizon''  ($k \gg |aH|$) during the bouncing phase,  a
modification of the dynamical evolution of the functions $C_2(k)$ and $C_4(k)$ would
occur. However, if the duration of the bouncing phase is short enough then one could
intuitively consider that the spectrum is left unchanged (although counterexamples exist
in the literature \cite{jmartin2003}). Therefore, one could perform a full analysis
regarding the CSL model during the bounce within the QMCU. Nonetheless, we will take a
pragmatical approach and assume that the shape of the spectrum, provided by the
functions $C_2(k)$ and $C_4(k)$, survives the bouncing phase and, then, we will use the
observational data to further constraint or completely discard the predicted spectra. In
case that the predicted spectra are consistent with observations, one can proceed to
perform the full-fledged analysis of implementing the CSL model to the QMCU during the
bouncing phase and study the possible corrections that may arise from passing the
perturbations through the bounce. This subject, however, will not be explored in the
present paper.

In the next section, we will explore the implications of the predicted spectra using the
observational data.

\section{Effects on the CMB temperature spectrum and its implications on the cosmological
parameters}
\label{effectsCMB}

The aim of this section is to analyze the viability of the CSL model by comparing the
corresponding predictions with the ones coming from the best fit canonical model to the CMB data. In
particular, we will focus on the power spectra obtained using the CSL model and its effect on the  angular power spectrum.

In order to perform our analysis, we start by setting the cosmological parameters of our
fiducial model, which will be used as a reference to compare with the CSL inspired spectra.
The fiducial cosmology will be the best fitting flat $\Lambda$CDM model from Planck data, with the following cosmological
parameters and values: baryon density in units of the critical density $\Omega_{\textrm{B}} h^2
=0.0223 $, dark matter density in units of the critical density $\Omega_{\textrm{CDM}}
h^2 = 0.1188$, Hubble constant $H_0 = 67.74$ in units of km s$^{-1}$ Mpc$^{-1}$,
reionization optical depth $\mathcal{T} = 0.066$, the scalar spectral index $n_s=
0.96$ and a pivot scale of $k_0 = 0.05$ Mpc$^{-1}$. These values can be found in the Table 4 presented by the latest Planck Collaboration
\cite{planckcmb2}.

Furthermore, we recall that the primordial power spectrum and the angular power spectrum
are related by \eq{Clstd}, i.e.
\beq
C_l = \frac{4 \pi}{25} \int \frac{dk}{k} \:   \Delta_l^2 (k) \mP_s(k).
\eeq
Hence, we will use the CSL predicted power spectra $\mP_s(k) = A_s (k/k_0)^{n_s-1}
C_i(k)$ with $i=1,2,3,4$, which correspond to the four different cases that we have considered so
far. We will focus first on the inflationary model of the early Universe and, then, on the
QMCU model. Also, note that the fiducial model corresponds to: $\mP_s(k) = A_s (k/k_0)^{n_s-1}$.
The precise prediction for the angular power spectrum will be obtained by using
the Boltzmann code CAMB \cite{CAMB}, with the aforementioned cosmological parameters.

\subsection{The angular power spectrum and the CSL model during inflation}

During inflation, the CSL power spectra is characterized by the functions $C_1(k)$
and $C_3(k)$, \eqs{C1} and \eqref{C3}, with standard spectral index $n_s -1 = -4 \epsilon
+ 2\delta$ and amplitude
\beq\label{ampinfcsl}
A_s = \left( \frac{V \lambda |\tau|}{12 \pi^2 M_P^4 \epsilon} \right)\bigg|_{k_0=aH}
\eeq

The output of the CAMB code, that is, the temperature autocorrelation power
spectrum of the fiducial model and the one provided by the CSL model during inflation are
indistinguishable; thus, we have decided not to show the plots. Instead, we present the
the relative difference, which we define as
\beq\label{Sl}
S(l) \equiv \frac{|C_l^{\textrm{CSL}} - C_l^{\textrm{fiducial}}
  |}{C_l^{\textrm{fiducial}}}
\eeq

\begin{figure}
\begin{center}
\includegraphics[scale=0.85]{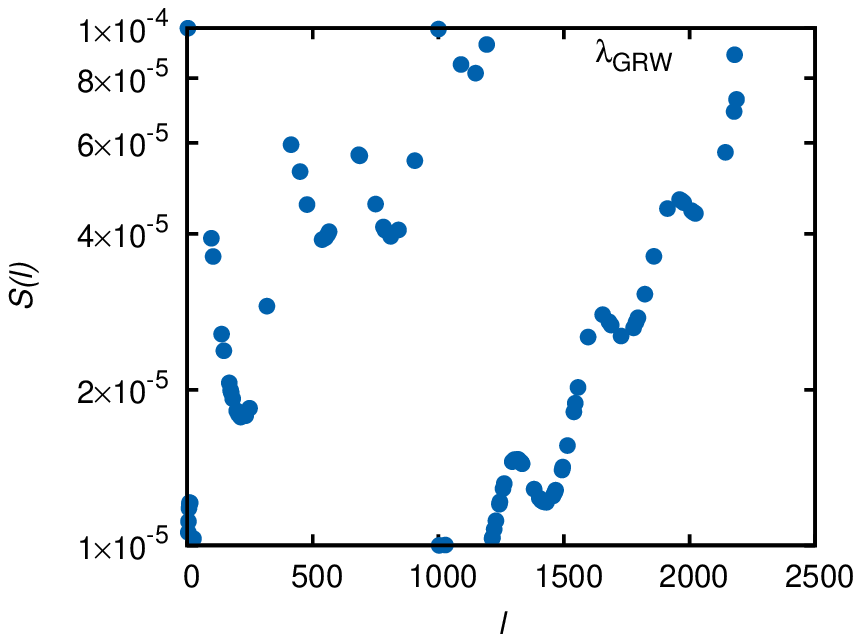}
\includegraphics[scale=0.85]{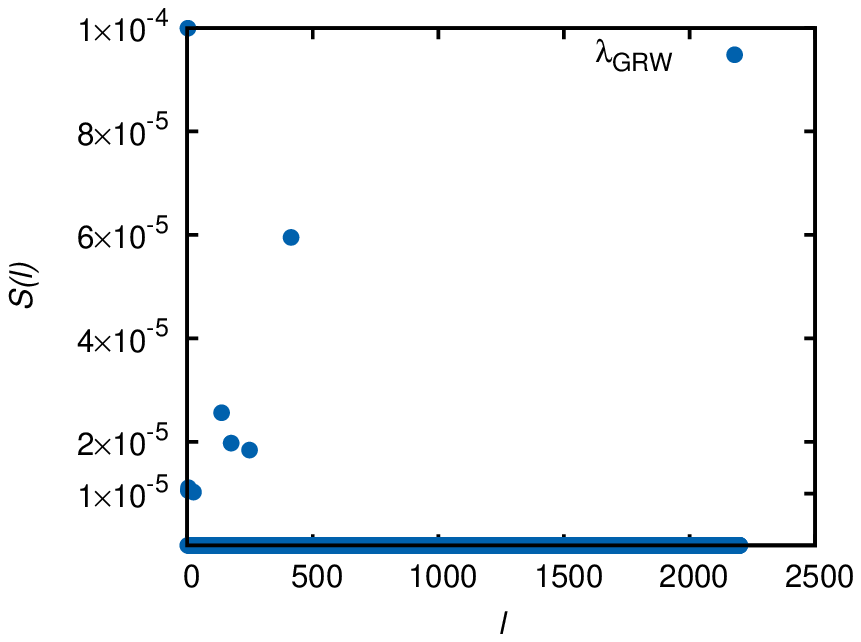}
\end{center}
\caption{The relative difference $S(l)$ [defined in Eq. \eqref{Sl}] between the angular
power spectrum predicted using the fiducial model and the one provided by the CSL model
during inflation.
Left: $\hat v_{\nk}^{R,I}$ as the collapse operator.
Right: $\hat p_{\nk}^{R,I}$ as the collapse operator.
For the CSL model, we have chosen the value of
$\lambda_{\textrm{GRW}} = 1.029 \times 10^{-2}$ Mpc$^{-1}$. Also, we have assumed
that inflation ends at an energy scale of $10^{-5} M_P$ and a total amount of inflation
corresponding to $N=65$, i.e. $\tau = 7.8 \times 10^{7}$ Mpc and $\eta_f = 4.6 \times
10^{-21}$ Mpc. Other values of $\lambda$ were plotted (not shown) achieving an excellent
fit, but they are observationally ruled out by their $\epsilon$ and $r$ values predicted.
See Table \ref{tabla1} and text for details.}
\label{Clinf}
\end{figure}

Figure \ref{Clinf} shows the relative difference between both
predictions.  On the left, we have chosen
$\hat v_{\nk}^{R,I}$ as the collapse operator
while on the right we have chosen $\hat p_{\nk}^{R,I}$.
In both cases, we have set an energy scale of $10^{-5} M_P$ for the energy at which
inflation ends, and a total amount of inflation corresponding to 65 $e$-foldings.

We observe that the relative difference between the fiducial spectrum and the one
predicted using the CSL model with, for instance $\lambda_{\textrm{GRW}}$, is practically null (the
highest difference is around $0.01\%$). This statement applies to both elections of the collapse
operator and for other $\lambda$ values listed in Table \ref{tabla1} (not shown in the figure).

We have also checked that the essentially null relative difference between the
fiducial model and the CSL model during inflation is also present in the $E$ polarization
autocorrelation power spectrum $C_l^{\textrm{EE}}$ and the temperature polarization
cross correlation power spectrum $C_l^{\textrm{TE}}$.

On the other hand, the amplitude of the power spectrum $A_s$ consistent with the CMB data
is $A_s \simeq
10^{-9}$ \cite{planckcmb2}. Henceforth, the amplitude obtained using the CSL model, as shown in
\eq{ampinfcsl}, must satisfy
\beq
10^{-9} \simeq \left( \frac{V \lambda |\tau|}{12 \pi^2 M_P^4 \epsilon} \right)
\bigg|_{k_0=aH}
\eeq
Clearly, different values of $\lambda$ will have an effect on the amplitude of the spectrum.

Assuming that the pivot scale $k_0$ crosses the Hubble radius at an energy scale of
$V_0^{1/4} = 10^{-4} M_P$ (i.e. one order of magnitude less than the presumed energy
at which inflation ends), an estimate for $\epsilon$ can be calculated. Therefore, the above
equation leads to
\beq\label{epsilonest}
\epsilon \simeq \frac{ \lambda |\tau| 10^{-7}}{12 \pi^2}.
\eeq

Table \ref{tabla1} shows the different values of $\epsilon$ obtained by considering several $\lambda$ values.
Also, in the same table, we provide an estimate for the tensor-to-scalar ratio $r$ (recall that the CSL model
predicts the same relation as standard inflation, i.e $r=16 \epsilon$). From Table
\ref{tabla1}, it can be seen that only the value corresponding to $\lambda_{\textrm{GRW}}$ is
consistent with both, the observed shape and amplitude of the spectrum. In
particular, assuming a characteristic energy scale of inflation of $10^{-4} M_P \simeq
10^{14}$ GeV, a total amount of inflation corresponding to $N=65$, and the value of
$\lambda_{\textrm{GRW}}$, we obtain an angular spectrum with a shape and an amplitude
that is indistinguishable from the fiducial model, which we know is
consistent with the observational data. The amplitude of the spectrum for this
particular set of values leads to an estimate for the slow roll parameter and the tensor-to-scalar
ratio of $\epsilon \simeq 10^{-4}$ and $r\simeq 10^{-2}$, respectively. Those values of
$\epsilon$ and $r$ are consistent with the ones presented by the latest results of the
Planck Collaboration \cite{Planckinflation15}.

\begin{table}[]
\centering
\caption{Estimation of $\epsilon$ from \eq{epsilonest}. We have used $|\tau| = 7.8 \times
10^{7}$ Mpc, which corresponds to $V^{1/4}=10^{14}$ GeV and $N=65$,  and the four
values of $\lambda$ shown below. Also, we have estimated the tensor-to-scalar ratio using
$r=16 \epsilon$. Note that only the $\lambda_{\textrm{GRW}}$ case is compatible with the
latest observations from Planck Collaboration \cite{planckcmb2}.}
\label{tabla1}
\begin{tabular}{|l|l|l|l|l|}
\hline
\multicolumn{1}{|c|}{$\lambda$ type} & \multicolumn{1}{c|}{$\lambda$ {[}s$^{-1}${]}} &
\multicolumn{1}{c|}{$\lambda$ {[}Mpc$^{-1}${]}} & \multicolumn{1}{c|}{$\epsilon$} &
\multicolumn{1}{c|}{$r$} \\ \hline
$\lambda_{\textrm{GRW}}$             & $10^{-16}$                                    &
$1.029 \times 10^{-2}$                          & $6.77 \times 10^{-4}$           & 0.01
                   \\
$\lambda_{1}$                        & $10^{-12}$                                    &
102.9                                           & 6.77 (not compatible)           & 108
(not compatible)      \\
$\lambda_{2}$                        & $10^{-10}$                                    &
10293                                           & 678  (not compatible)           & 10848
(not compatible)    \\
$\lambda_{\textrm{Adler}}$           & $10^{-8}$                                     &
1029378                                         & 67793 (not compatible)          &
1084688 (not compatible) \\ \hline
\end{tabular}
\end{table}

It is also instructive to mention that, if future observations confirm the results
of the BICEP2 Collaboration \cite{BICEP2}, i.e. $r \simeq 0.2$, then the value of
$\lambda_{\textrm{GRW}}$ would not be compatible with the values of $V$ and $N$ used in Table
\ref{tabla1}. In fact, the $\lambda$ value that might be compatible would be one such that $\lambda
\ll \lambda_{\textrm{GRW}}$. That would open a new range of parameter space
to explore in addition to considering other experimental setups different from
cosmological ones \cite{bassibounds2016,bassibounds22016}.

\subsection{The angular power spectrum and the CSL model during the quasi-matter
contracting phase}

In the QMCU framework, the power spectra are characterized by the functions $C_2(k)$
and $C_4(k)$, i.e. \eqs{C2} and \eqref{C4}, respectively. The predicted scalar spectral index is
$n_s -1 = 12 \bepsilon$, and the amplitude is given by
\beq\label{ampqmatcsl}
A_s \simeq \frac{\rho_c}{\rho_P} \frac{\lambda |\tau|}{2}.
\eeq
Notice we have approximated the integral that appears in the amplitude, corresponding
to \eq{Aynscontraccion}, by $\rho_c/\rho_P$ [see Eqs. \eqref{Asoriginal1} and
\eqref{Asoriginal2}].

\begin{figure}
\begin{center}
\includegraphics[scale=0.85]{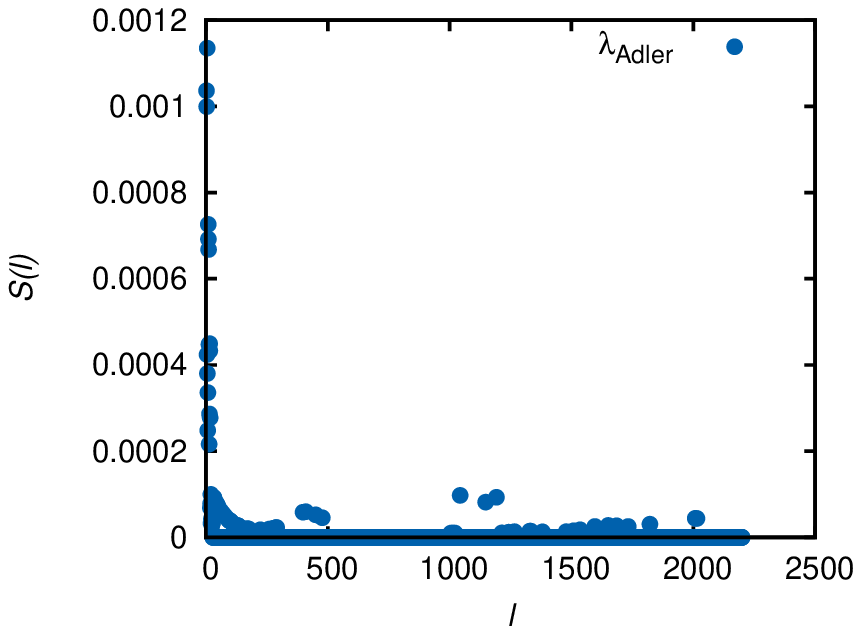}
\includegraphics[scale=0.85]{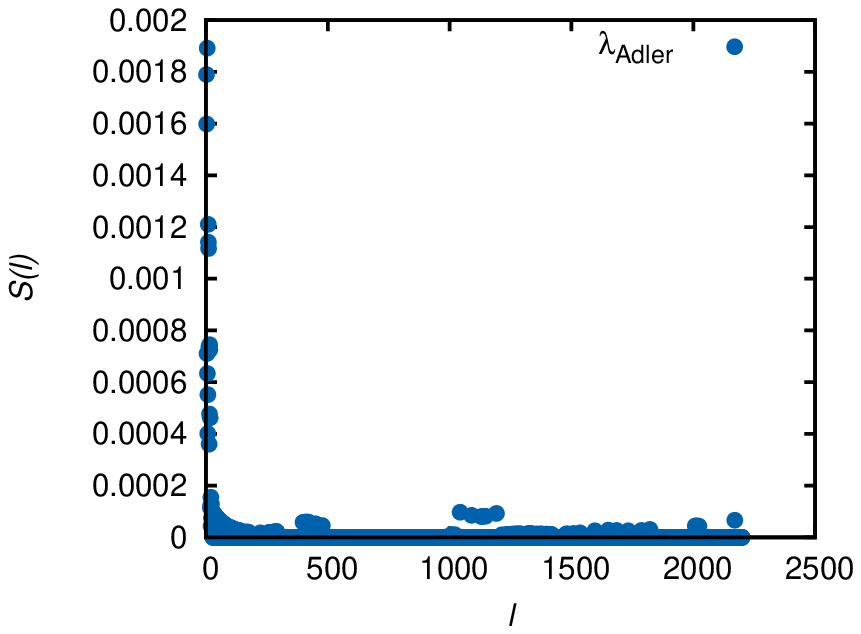}
\end{center}
\caption{The relative difference $S(l)$ [defined in Eq. \eqref{Sl}] between the angular
power spectrum predicted using the fiducial model and the one provided by the CSL model
during the quasi-matter contracting phase. Left: $\hat v_{\nk}^{R,I}$ as the collapse operator.
Right: $\hat p_{\nk}^{R,I}$ as the collapse operator. For the CSL model, we have chosen,
as an illustrative example, the value of
$\lambda_{\textrm{Adler}} = 1029378$ Mpc$^{-1}$. Also, we have assumed that
the quasi-matter contracting phase begins at conformal time $|\tau| = 1.15 \times
10^{8}$ Mpc and lasts a total amount of $N=50$ $e$-foldings, with $-\eta_f = 1.6 \times
10^{-3}$ Mpc. Other values of $\lambda$ listed in Table \ref{tabla2} also achieve an
excellent fit (not shown), but the analysis done in this work does not allow prefer one
value over another.}
\label{Clqmat}
\end{figure}

The output of the CAMB code, that is, the temperature autocorrelation power
spectrum of the fiducial model and the one provided by the CSL model during the QMCU are
also indistinguishable. Thus, we present again the relative difference defined in
\eq{Sl} where now $C_l^{\textrm{CSL}}$ corresponds to the angular power spectrum during
the QMCU.

Figure \ref{Clqmat} shows the relative difference between both
predictions.  On the left, we have chosen $\hat v_{\nk}^{R,I}$ as the collapse operator
while on the right we have chosen $\hat p_{\nk}^{R,I}$.
In both cases, we have assumed a total duration of $50$ $e$-foldings for the
quasi-matter contracting phase and a conformal time
$|\tau| \simeq 10^8$ Mpc corresponding to the beginning of the contracting stage.
We show only the plot of $S(l)$ corresponding to the value of $\lambda_{\textrm{Adler}}$
merely as an illustrative example; the plots for the values corresponding to
$\lambda_1$, $\lambda_2$, $\lambda_{\textrm{GRW}}$ follow the exactly same behavior as
the one shown in Fig. \ref{Clqmat}.

We found no difference between the fiducial spectrum and the one
provided by the CSL model for the four values of $\lambda$ listed in Table \ref{tabla2};
the highest relative difference is around $0.1\%$. This statement
applies to both elections of the collapse operator. Finally, we have also checked that
the essentially null relative difference between the
fiducial model and the CSL model during the QMCU is also present in the $E$ polarization
autocorrelation power spectrum $C_l^{\textrm{EE}}$ and the temperature polarization
cross correlation power spectrum $C_l^{\textrm{TE}}$.

On the other hand, the amplitude of the power spectrum $A_s$ consistent with the CMB data is $A_s \simeq
10^{-9}$ \cite{planckcmb2}. Therefore, the ratio $\rho_c/\rho_P$, obtained using the CSL model,
as shown in \eq{ampqmatcsl}, must satisfy
\beq\label{defkappa1}
\frac{\rho_c}{\rho_P} \simeq \kappa_1 \qquad \textrm{where} \qquad \kappa_1 \equiv
\frac{2}{\lambda |\tau|} \times 10^{-9}.
\eeq
Consequently, an estimate for the energy scale of the critical energy $E_c$ is
\beq\label{defkappa2}
E_c \simeq 10^{\kappa_2} M_P \qquad \textrm{where} \qquad \kappa_2 \equiv \frac{1}{4}
\log \kappa_1.
\eeq

In Table \ref{tabla2}, we show the different values of $\kappa_1$ and $\kappa_2$ by using the
chosen $\lambda$ values, with $|\tau| = 1.15 \times10^{8}$ Mpc. We infer that the
four values of $\lambda$ considered are consistent with a critical energy scale in the
range $(10^{-3} M_P, 10^{-6} M_P)$.

\begin{table}[]
\centering
\caption{Estimation of $\kappa_1$ and $\kappa_2$ defined in \eqs{defkappa1} and
  \eqref{defkappa2} respectively. We have used $|\tau| \simeq 10^{8}$ Mpc and the four
values of $\lambda$ shown below. Also we have estimated
the quantities $\lambda |\tau|$.}
\label{tabla2}
\begin{tabular}{|l|l|l|l|l|l|}
\hline
\multicolumn{1}{|c|}{$\lambda$ type} & \multicolumn{1}{c|}{$\lambda$ {[}s$^{-1}${]}} &
\multicolumn{1}{c|}{$\lambda$ {[}Mpc$^{-1}${]}} & \multicolumn{1}{c|}{$\lambda |\tau|$} &
\multicolumn{1}{c|}{$\kappa_1$} & $ \kappa_2$ \\ \hline
$\lambda_{\textrm{GRW}}$             & $10^{-16}$                                    &
$1.029 \times 10^{-2}$                          & $1.029 \times 10^{6}$                 &
$10^{-15}$                    & -3.75                     \\
$\lambda_{1}$                        & $10^{-12}$                                    &
102.9                                           & $1.029 \times 10^{10}$                &
$10^{-19}$                    & -4.75                     \\
$\lambda_{2}$                        & $10^{-10}$                                    &
10293                                           & $1.029 \times 10^{12}$                &
$10^{-21}$                    & -5.25                     \\
$\lambda_{\textrm{Adler}}$           & $10^{-8}$                                     &
1029378                                         & $1.029 \times 10^{14}$                &
$10^{-23}$                    & -5.75         \\ \hline
\end{tabular}
\end{table}

It is worthwhile to mention that, in the QMCU, the spectral index $n_s$ and the
tensor-to-scalar ratio $r$ are not related each other as in the standard inflationary
paradigm [see \eqs{rQMCU} and \eqref{Aynscontraccion}]. However, the spectral index, along with the running of
the spectral index $\alpha_s \equiv d n_s/d \ln k$, are the two main
parameters of the QMCU model used to compare with the observational data
\cite{elizalde,jaume,amoros}. The CSL model applied to the QMCU does not affect
those parameters. The only observable affected by the CSL model is the amplitude of the spectrum
(which is not related to the parameter $r$ as in the
standard spectrum). Consequently, in order to put an upper bound to the energy scale at
which the bouncing phase begins, and which would be equivalent to set a constraint on the
parameter $\lambda$, one should consider a specific theoretical model of the QMCU (i.e.
to choose a specific dynamics and a potential of the field $\varphi$).
Therefore, in the QMCU, and with the same degree of accuracy of past works dealing with the same model,
all of the four values of the CSL parameter $\lambda$ considered
here yield consistent predictions with observational data; specifically, the predictions
regarding the shape of the spectrum, the scalar spectral index and the running of the spectral
index.

On the other hand, note that the information that could discriminate among different values of
$\lambda$ is codified in the amplitude of the spectrum. The predicted amplitude of the
spectrum depends on the critical energy density $\rho_c$ (the value of the energy density at the
bouncing time), which is model dependent.

\section{Conclusions}
\label{conclusions}

The CSL model is a physical mechanism that attempts to provide a solution to the
measurement problem of Quantum Mechanics by modifying the Schr\"{o}dinger equation. The
CSL model can be referred as an objective reduction mechanism or ``effective collapse''
of the wave function, and one of the main elements of this model is the collapse operator,
i.e. the operator whose eigenstates correspond to the evolved states by the collapse
mechanism. Also, in principle, it is possible to apply such a mechanism to any physical system.

In this work, we have applied the CSL model to the early Universe by considering two
cosmological models: the matter bounce scenario (MBS) and standard slow roll inflation.
Additionally, we have considered two different collapse schemes, one
in which the field variable (given in terms of the Mukhanov-Sasaki variable) serves as
the collapse operator, and other scheme where the collapse operator is the conjugated
momentum.

In all cases, we have found a prediction for the primordial power spectrum, which is a
function of the standard parameters of each cosmological model, and also of the CSL
parameter $\lambda$. Although the exact expressions for the primordial power spectra are
different in each case, there are features that are essentially the same as its
standard--non-collapse--counterparts. Specifically, the predictions for the scalar
spectral index and the tensor-to-scalar ratio are exactly the same as the ones given in the
MBS and slow roll inflation without collapse. On the other hand, in each case, the shape
of the spectrum is modified by a function of the wave number $k$, associated to the modes
of the field, and by the inclusion of the $\lambda$ parameter.
However, for a suitable choice of values corresponding to the parameters of the
cosmological models, there is no significant change in the prediction for the CMB angular
power spectrum (i.e. the $C_l$'s) that can be distinguished from the canonical flat $\Lambda$CDM model.

Meanwhile, the prediction for the amplitude of the spectrum is modified directly by the
parameter $\lambda$. We have empirically explored the range of values of $\lambda$, from
the originally value suggested by Ghirardi-Rimini-Weber (GRW) $\lambda_{\textrm{GRW}}
\simeq 10^{-16}$ s$^{-1}$ \cite{GRW}, to the one given by Adler $\lambda_{\textrm{Adler}} \simeq
10^{-8}$ s$^{-1}$ \cite{adler2}. In the case of slow roll inflation, we have found that for a
characteristic energy scale of $10^{14}$ GeV and a total amount of inflation of 65
$e$-folds, only the value suggested by GRW is compatible with the observational bound of
the amplitude; other values of $\lambda$ greater than $\lambda_{\textrm{GRW}}$, e.g.
$\lambda_{\textrm{Adler}}$,  cannot be made compatible with the observed amplitude
(because that would require values for the slow roll parameter such that $\epsilon > 1$).
In the MBS case, we have found that the modification in the predicted amplitude of
the spectrum, given by the $\lambda$ parameter, causes that the critical energy density
$\rho_c$, i.e. the energy density at which the bouncing phase begins, to be several orders
of magnitude less than the Planck energy density $\rho_P$. The precise number of orders
of magnitude varies according to the value of $\lambda$. For instance, by assuming
$\lambda_{\textrm{GRW}}$
and a total amount of $\sim 50$ $e$-folds for the matter dominated contracting phase, we have $\rho_c
\simeq 10^{-15} \rho_P$. The latter relation is obtained by requiring the compatibility
between the predicted amplitude of the scalar power spectrum and the one from the Planck CMB data $A_s \sim 10^{-9}$.

In conclusion, it was possible to incorporate the CSL model into the cosmological context again, in
particular when dealing with the quantum-to-classical transition of the primordial
inhomogeneities. Moreover, it is remarkable that our implementation of the CSL model yields predictions that
are also in agreement with experiments in the regimes so far investigated empirically. Those
experiments involve values of the CSL parameter $\lambda$ that have been tested in
laboratory settings, quite disengaged from the cosmological framework. We acknowledge
that at this stage, the application of CSL
model to the early Universe, as done in this manuscript, can be seen as an \emph{ad hoc} employment. However, the fact that the predictions can be empirically tested make us hopeful that future studies will overcome the perceived shortcomings.

\acknowledgements
G. L.'s research was funded by Consejo Nacional de Ciencia y Tecnolog\'ia, CONACYT
(Mexico). G. R. B. acknowledges support from CONICET (Argentina) Grant PIP 112-2012-0100540. S. J. L. and G. L. are supported by PIP 11220120100504 CONICET (Argentina).

\appendix

\section{Calculations of Section \ref{SecImplementation} }\label{app}

In this Appendix, we provide a sketch of the computational steps that led to the
results presented in Sect. \ref{SecImplementation}. In the first half of this Appendix, we will
focus on cases (i) and (iii), which correspond to selecting the MS variable as the
collapse operator during inflation and the QMCU, respectively. The second half will contain
the details according to cases (ii) and (iv), corresponding to selecting $\hat
p_{\nk}^{R,I}$ as the collapse operator during inflation and the QMCU, respectively.

\subsection{The field $\hat v_{\nk}$ as the collapse operator}

We will proceed the calculation of the scalar power spectrum by dealing simultaneously with the inflationary
and the QMCU frameworks; and finally, we will argue that the computation is similar for the tensor
power spectrum.

In order to obtain the scalar power spectrum, we will use Eq. \eqref{restav}. Let us focus
first on the second term on the right hand side in that equation, which can be
obtained by using the CSL evolution equation \eqref{cslmodos}. Recall that, since in this
case $\hat \varTheta_{\nk} = \hat v_{\nk}$, it will be convenient to work with the wave
function in the field representation Eq. \eqref{psiondav}. The CSL evolution equation leads to the following equation of motion:
\beq
A_k' = \frac{ik^2}{2} + \lambda_k - \frac{2 \beta}{\eta} A_k - 2i A_k^2.
\eeq

The previous equation is solved by performing the change of variable
$A_k(\eta) \equiv f_k'(\eta)/[2i f_k(\eta)]$, resulting in a Bessel differential
equation for $f_k(\eta)$. After solving such an equation, and returning to the original
variable $A_k$, we obtain
\beq\label{Aexacta}
A_k(\eta) = \frac{q}{2i} \left[\frac{-J_{m-1} (-q\eta)-e^{-i\pi m}
J_{1-m} (-q\eta)}{J_{m} (-q\eta) -e^{-i\pi m} J_{-m}
(-q\eta)}
      \right],
\eeq
being $q^2 \equiv k^2 (1-2i\lambda_k/k^2)$, and where the initial condition for the
Bunch-Davies vacuum $A_k(\tau ) = k/2$ was used (we remind the reader that $\tau$
corresponds to the conformal time at the beginning of the contraction phase or the
exponential expansion). The function $J_m$ is a Bessel function of the first
kind of order $m \equiv 1/2 - \beta$ [recall that $\beta$ is defined in Eq.
\eqref{zbeta}].

We now expand $A_k(\eta)$ in the limit where the proper wavelength associated to
the modes becomes larger than the Hubble radius $H^{-1}$, that is, in the limit $-q\eta
\to 0$ (provided that $\lambda_k \ll 1$). However, note that $\beta$ varies according whether
inflation or a contracting phase is assumed. For inflation $\beta \simeq
-1$ whilst for the QMCU case $\beta \simeq 2$. This implies that the dominant modes
correspond to $m \simeq 3/2$ during inflation. On the contrary, in the QMCU, the
dominant modes correspond to $m \simeq -3/2$. Furthermore, we are only
interested in the expansion on the real part of $A_k (\eta)$. Therefore, for case (i)
the expansion of the real part of $A_k(\eta)$ is
\beq\label{reacaso1}
\text{Re} \: A_k(\eta) \simeq \frac{\lambda_k}{2k(m-1)}(-k\eta) + \zeta_k^{2m}
\frac{\sin(\pi m+ 2m\theta_k) }{\sin (\pi m)} \frac{k \pi}{2^{2m} \Gamma(m)^2}
(-k\eta)^{2m-1},
\eeq
and for case (ii) the corresponding expansion is
\beq\label{reacaso2}
\text{Re}\:  A_k(\eta) \simeq \frac{\lambda_k}{2k(-m-1)}(-k\eta) + \zeta_k^{-2m}
\frac{\sin(\pi m+ 2m\theta_k) }{\sin (\pi m)} \frac{k \pi}{2^{-2m} \Gamma(-m)^2}
(-k\eta)^{-2m-1}.
\eeq
In both cases, $\zeta_k$ and $\theta_k$ are given in \eq{zetakv}.

Next, we focus on the first term of \eq{restav}, i.e.  $\overline{\bra \hat
v_{\nk}^2 \ket}$. It will be useful to define the following quantities:
\beq\label{defQRS}
Q \equiv \overline{\bra \hat v_{\nk}^2 \ket}, R\equiv \overline{\bra \hat p_{\nk}^2
  \ket} \:\: \textrm{and} \: \: S \equiv \overline{\bra \hat p_{\nk} \hat v_{\nk} +
\hat v_{\nk} \hat p_{\nk} \ket}.
\eeq
The equations of evolution for $Q,R$ and $S$ are obtained using \eq{operadorcslmodos}, with $\hat \varTheta_{\nk} = \hat v_{\nk}$. That is,
\beq\label{QRS}
Q' = S + \frac{2\beta}{\eta}  Q, \qquad R' = -k^2 S - \frac{2 \beta}{\eta}  R +
\lambda_k, \qquad S' = 2R - 2k^2 Q.
 \eeq

Therefore, we have a linear system of coupled differential equations, whose
general solution is a particular solution to the system plus a solution to the
homogeneous equation (with $\lambda_k=0$). After a long series of calculations
we find:
\beq\label{Q}
 Q(\eta) = (-k\eta) [ C_1J_n^2(-k\eta)  + C_2 J_{-n}^2(-k\eta)+ C_3
 J_n(-k\eta) J_{-n}(-k\eta)] + \frac{ \lambda_k \eta}{2k^2 },
 \eeq
with
\begin{eqnarray}\label{nyalpha}
  n = \frac{3}{2} + \frac{2}{3} \alpha;  \qquad \textrm{and} \qquad \alpha
  \equiv \begin{cases} 3 \epsilon - \frac{3}{2} \delta
\:\: \textrm{if assuming the inflationary Universe }\\
-9 \bar \epsilon \:\: \textrm{if assuming the QMCU }  \\
\end{cases}
\end{eqnarray}
and the constants $C_1, C_2$ and $C_3$ are found by imposing the initial
conditions corresponding to the Bunch-Davies vacuum state: $Q(\tau)=1/(2k),
R(\tau) = k/2$ and $S(\tau)=0$. Equation \eqref{Q} is exact; expanding it again
around $-k\eta \to0$ yields
\beq\label{Qaprox}
Q(\eta) \simeq \frac{\pi}{2k^2 \sin^2(n\pi)} \bigg\{  \frac{k}{2} - \frac{\lambda_k
  \tau}{2} + \frac{m \lambda_k}{k} \sin\Delta \cos\Delta     \bigg\}
 \frac{2^{2n}}{\Gamma^2(1-n)} (-k\eta)^{-2n+1},
\eeq
where
\beq
\Delta = -k\tau -\frac{n \pi}{2} -\frac{\pi}{4}.
\eeq

Using the above results, we can compute the quantity  $\overline{\bra \hat v_{\nk}
\ket^2}$ using Eq. \eqref{restav}. In case (i), we substitute Eqs. \eqref{reacaso1} and
\eqref{Qaprox} into Eq. \eqref{restav}, obtaining
\barr\label{eqchingona3}
\overline{\bra \hat v_{\nk} \ket^2} &=& Q(\eta) - \frac{1}{4 \textrm{Re} A_k (\eta)} \nn
&\simeq& \frac{\pi}{2k^2 \sin^2(n\pi)} \bigg\{  \frac{k}{2} - \frac{\lambda_k
  \tau}{2} + \frac{m \lambda_k}{k} \sin\Delta \cos\Delta    \bigg\}
 \frac{2^{2n}}{\Gamma^2(1-n)} (-k\eta)^{-2n+1} \nn
 &-& \frac{1}{4} \left[ \frac{\lambda_k}{2k(m-1)}(-k\eta) + \zeta_k^{2m}
\frac{\sin(\pi m+ 2m\theta_k) }{\sin (\pi m)} \frac{k \pi}{2^{2m} \Gamma(m)^2}
(-k\eta)^{2m-1} \right]^{-1}.
\earr

With the expression in \eqref{eqchingona3} at hand (which is valid for
$\overline{\bra \hat v_{\nk}^{R}
\ket^2}$ and  $\overline{\bra \hat v_{\nk}^{I} \ket^2}$), and using Eq. \eqref{PSchingon},
our predicted scalar power spectrum during inflation (at the lowest order in the slow roll parameter) is given in \eq{PSescalarcaso1}.
(we have also used that during inflation $z^2(\eta) \simeq
2 \epsilon M_P^2/ (H^2 \eta^2)$ and $ m = n  = 3/2 + 2 \epsilon - \delta$).

Analogously, in case (ii), substituting Eqs. \eqref{reacaso2} and
\eqref{Qaprox} into \eqref{restav} yields
\barr\label{eqchingona4}
\overline{\bra \hat v_{\nk} \ket^2} &=& Q(\eta) - \frac{1}{4 \textrm{Re} A_k (\eta)} \nn
&\simeq& \frac{\pi}{2k^2 \sin^2(n\pi)} \bigg\{  \frac{k}{2} - \frac{\lambda_k
  \tau}{2} + \frac{m \lambda_k}{k} \sin\Delta \cos\Delta     \bigg\}
 \frac{2^{2n}}{\Gamma^2(1-n)} (-k\eta)^{-2n+1}\nn
 &-& \frac{1}{4} \left[  \frac{\lambda_k}{2k(-m-1)}(-k\eta) + \zeta_k^{-2m}
\frac{\sin(\pi m+ 2m\theta_k) }{\sin (\pi m)} \frac{k \pi}{2^{-2m} \Gamma(-m)^2}
(-k\eta)^{-2m-1}   \right]^{-1}.
\earr

Hence, substituting the above expression in Eq. \eqref{PSchingon} yields:
\barr\label{PStmp}
P_s(k) &=& \frac{(-k\eta)^{-2\mu_s+3}}{4\pi^2z^2 \eta^2} \bigg\{
\frac{2^{2\mu_s} \pi}{\sin^2(\mu_s \pi) \Gamma^2(1-\mu_s)}   \bigg[ 1- \frac{\lambda_k
  \tau}{k} - \frac{3 \lambda_k}{k^2} \sin (-k\tau) \cos (-k\tau)  \bigg]    \nn
&-&  \left[ \frac{\lambda_k}{2(\mu_s-1)k^2} (-k\eta)^{-2\mu_s+2} +
\frac{\zeta_k^{2\mu_s}\pi \sin(\pi \mu_s + 2 \mu_s \theta_k)    }{\sin (\pi \mu_s)
2^{2\mu_s}\Gamma^2 (\mu_s) }        \right]^{-1} \bigg\}, \qquad \mu_s \equiv
\frac{3}{2}
- 6 \bar \epsilon
\earr
where we have used $m=-n = 3/2 - 6 \bar \epsilon$. As argued in Refs.
\cite{elizalde,jaume}, during the quasi-matter contracting phase $z \simeq \eta^2/(3
\sqrt{3})$ and $|aH| = -2/\eta$, which implies that Eq. \eqref{PStmp} can be written as
in the final form of the power spectrum presented in \eq{PSescalarcaso2}

Let us focus now on the tensor modes. The action for the tensor perturbations is obtained
from the Einstein-Hilbert action by expanding the tensor perturbations $h_{ij} (\x,\eta)$ up to
second-order \cite{mukhanov1992}. The resulting action for the tensor field
$h_{ij}
(\x,\eta)$ can be expressed in terms of its Fourier modes $h_{ij}
(\nk,\eta)=h_{\nk} (\eta) e_{ij} (\nk)$, with $e_{ij} (\nk)$ representing a
time-independent polarization tensor. Performing the change of variable
\begin{equation}\label{cambiovariable}
h_{\nk} (\eta) \equiv \frac{2}{M_P (e^i_{\,j}e^j_{\,i})^{1/2}} \frac{v_{\nk}
(\eta)}{a(\eta)},
\end{equation}
the action can be written as $\delta^{(2)} S_h =\frac{1}{2} \int d\eta \:d^3
\nk \:\mathcal{L}_h$, where
\beq\label{accionh}
\mathcal{L}_h =  v_{\nk}' v_{\nk}^{\star '} - k^2 v_{\nk} v_{\nk}^{\star}
-\frac{a'}{a}  \left(  v_{\nk} v_{\nk}^{\star '}  +  v_{\nk}'  v_{\nk}^{\star}
\right) +  \left( \frac{a'}{a} \right)^2 v_{\nk} v_{\nk}^{\star}.
\eeq
The Lagrangian in \eq{accionv}, and the one in \eq{accionh}, share the same structure. In particular, if one replaces $z'/z \to a'/a$ in \eq{accionv},
one obtains \eq{accionh}. Thus, the preceding calculations can be directly employed to obtain the tensor power spectrum by replacing $z'/z \to a'/a$ in the whole procedure. The explicit form of this last quantity is
\begin{eqnarray}\label{abeta}
  \frac{a'(\eta)}{a(\eta)} = \frac{\tilde \beta}{\eta};  \qquad \textrm{where} \qquad
\tilde \beta
\equiv \begin{cases} -(1+\epsilon)
\:\: \textrm{if assuming the inflationary Universe }\\
2(1-3\bepsilon) \:\: \textrm{if assuming the QMCU } . \\
\end{cases}
\end{eqnarray}
Consequently, the replacement $\beta \to \tilde \beta$, in the equations of the present subsection, allows us to obtain the tensor power spectra
shown in Eqs. \eqref{PStensorcaso1} and \eqref{PStensorcaso2}.

\subsection{The momentum $\hat p_{\nk}$ as the collapse operator}

In the rest of this Appendix, we will present the mathematical details for obtaining the
scalar and tensor power spectra, using the CSL model, when the collapse operator is $\hat p_{\nk}$.
As in the previous subsection, we will use the framework provided by the inflationary
Universe and the QMCU simultaneously. This will complete the computation of the power
spectra for the last two cases mentioned at the beginning of Sect.
\ref{SecImplementation}, i.e. cases (iii) and (iv).

Since in this subsection we are considering that the collapse operator is the momentum
operator, $\hat \varTheta_{\nk} = \hat p_{\nk}^{R,I}$, it is convenient to work with the
wave function in the momentum representation, \eq{psiondap}. Moreover, as argued at the
beginning of this section, in that representation, the quantities of interest, namely
$\overline{\bra \hat v_{\nk}^{R}\ket^2}$ and  $\overline{\bra \hat v_{\nk}^{I} \ket^2}$,
can be calculated using \eq{restap}. Furthermore, in spite of the collapse operator being
the momentum operator, the calculations of the previous subsection serve as a blueprint
for the computations in this subsection. Once again, we will proceed by omitting the
indexes $R$,$I$.

Let us focus first on the second term of the right hand side of \eq{restap}, i.e.
$|\tilde A_k(\eta)|^2/\textrm{Re}[ \tilde A_k(\eta)]$. The motion equation for $\tilde
A_k(\eta)$ is obtained from the CSL \eq{cslmodos}, which leads to
\beq
\tilde A_k'(\eta) = \frac{i}{2} + \lambda_k + \frac{ 2\beta}{\eta} \tilde  A_k - 2ik^2
\tilde{A}_k(\eta)^2.
\eeq
Performing the change of variable $\tilde A_k \equiv g_k'(\eta)/[2ik^2g_k(\eta)]$ in the
last equation, results in a Bessel differential equation for $g_k(\eta)$. Now, using such a
solution and returning to the original variable, we have
\beq\label{A2}
\tilde A_k(\eta) = \frac{q}{2ik^2} \left[ \frac{ J_{m+1}(-q\eta) + e^{-im\pi} J_{-m-1}
(-q\eta)}{J_m(-q\eta) - e^{-im\pi} J_{-m} (-q\eta) }    \right], \qquad \qquad m \equiv
-\frac{1}{2} - \beta,
\eeq
where $ q^2\equiv k^2 (1-2i\lambda_k)$. Also, we have used the
initial condition provided by the Bunch-Davies vacuum, which is $\tilde{A}_k(\tau ) = 1/2k$ with
$\tau \to -\infty$. Do not confuse the $m$ of \eq{A2} with the one of the previous subsection.

The next step is to perform the expansion for $-q\eta \to 0$.
Note that if we consider the inflationary Universe, then $\beta \simeq -1$, which means
that $m \simeq 1/2$. On the other hand, for the QMCU case $\beta \simeq 2$, which implies that
$m \simeq -5/2$ [note that $\beta$ is defined in \eq{zbeta}].

In other words, for case (iii), we have the following expansion
\beq\label{termino2caso3}
\frac{|\tilde{A}_k(\eta )|^2}{ \text{Re}[A_k (\eta)]} \simeq \frac{\sin (\pi m)
\Gamma^2(m+1)2^{2m} \tilde \zeta_k^{-2m} (-k\eta)^{-2m-1}}{k
\sin(2m \tilde \theta_k + \pi m) \pi}.
\eeq

And in case (iv), the corresponding expansion results,
\beq\label{termino2caso4}
\frac{|\tilde{A}_k(\eta )|^2}{ \text{Re}[A_k (\eta)]} \simeq \frac{c_1\tilde \zeta_k^2}{2k} \left[
\frac{1+c_2 2 \tilde \zeta_k^2 \cos(2\tilde \theta_k) (-k\eta)^2 + c_2 2 \tilde \zeta_k^2
\cos[2(m+1)\tilde \theta_k + \pi m]  (-k\eta)^{-2m-2}}{\sin (2 \tilde \theta_k)
(-k\eta)^{2m+2}  + c_2 \tilde \zeta_k^2 \sin (4\tilde \theta_k)  (-k\eta)^{2m+4} - c_3
\tilde \zeta_k^{-2m-2}  \sin[2m\tilde \theta_k + \pi m] } \right] (-k\eta)^{2m+3}
\eeq
where
\beq
c_1 \equiv \frac{1}{2(m+1)}, \qquad \qquad c_2 \equiv \frac{1}{2^2
(m+1)(m+2)}, \qquad \qquad c_3 \equiv \frac{2^{2m+2}\Gamma(m+2)
}{\Gamma(-m)}.
\eeq
In both cases, the definitions of the quantities $\tilde{\zeta_k}$ and $\tilde \theta_k$
are given in \eq{zetakp}.

Now, we have to obtain the first term of the right hand side of \eq{restap}, that is,
$\overline{\bra \hat v_{\nk}^2 \ket}$. We will employ the same procedure
as in the previous subsection. We use the previous definitions for the quantities $Q(\eta)$,$R(\eta)$ and
$S(\eta)$ \eq{defQRS} and \eq{operadorcslmodos} but taking into account that
$\varTheta_{\nk}^{R,I} = \hat p_{\nk}^{R,I}$. Thus, the evolution equations are:
\beq\label{QRSoperadorp}
Q' = S+\frac{2\beta}{\eta}Q+\lambda_k, \enskip R'=-k^2 S - \frac{2\beta}{\eta}R, \enskip
S'=2R - 2k^2 Q.
\eeq
Those equations are solved using the initial conditions provided by $Q(\tau) = 1/(2k)$,
$R(\tau) = k/2$ and $S(\tau) = 0$.

We are mainly interested in the solution for $Q(\eta) \equiv \overline{\bra \hat
v_{\nk}^2 \ket}$ [which is the first term on the right hand side of  \eq{restap}]. Then, performing the series
expansion to the lowest order around $-k\eta \to 0$ yields
\beq\label{Q2}
Q(\eta) \simeq \frac{\pi}{2k^2 \sin^2(n\pi)} \bigg\{  \frac{k}{2} - \frac{k^2\lambda_k
  \tau}{2} + {m k\lambda_k} \sin\Delta \cos\Delta     \bigg\}
 \frac{2^{2n}}{\Gamma^2(1-n)} (-k\eta)^{-2n+1}
\eeq
where $\Delta = -k\tau -\frac{n \pi}{2} -\frac{\pi}{4}$ and $n$ is exactly the same as
the one defined in \eq{nyalpha}.

We are now in position to compute the scalar power spectrum. For case (iii), substituting \eqs{termino2caso3} and \eqref{Q2} into
\eq{restap} yields
\barr\label{eqchingonap3}
\overline{\bra \hat v_{\nk} \ket^2} &=& Q(\eta) - \frac{|A_k(\eta)|^2}{\textrm{Re} A_k
(\eta)} \nn
&\simeq& \frac{\pi}{2k^2 \sin^2(n\pi)} \bigg\{  \frac{k}{2} - \frac{k^2\lambda_k
  \tau}{2} + {mk \lambda_k} \sin\Delta \cos\Delta    \bigg\}
 \frac{2^{2n}}{\Gamma^2(1-n)} (-k\eta)^{-2n+1} \nn
 &-& \frac{\sin (\pi m)
\Gamma^2(m+1)2^{2m} \zeta_k^{-2m} (-k\eta)^{-2m-1}}{k
\sin(2m \theta_k + \pi m) \pi}.
\earr
In addition, by noting that $m=1/2+2\epsilon-\delta $ and $n = 3/2 + 2\epsilon -
\delta $, we substitute \eq{eqchingonap3} in \eq{PSchingon}, which results in the final
expression for the power spectrum shown in \eq{PSescalarcaso3}.

For case (iv), we substitute \eq{termino2caso4} and \eq{Q2} into \eq{restap} which results in
\barr\label{eqchingonap4}
\overline{\bra \hat v_{\nk} \ket^2} &=& Q(\eta) - \frac{|A_k(\eta)|^2}{\textrm{Re} A_k
(\eta)} \nn
&\simeq& \frac{\pi}{2k^2 \sin^2(n\pi)} \bigg\{  \frac{k}{2} - \frac{k^2\lambda_k
  \tau}{2} + {mk \lambda_k} \sin\Delta \cos\Delta    \bigg\}
 \frac{2^{2n}}{\Gamma^2(1-n)} (-k\eta)^{-2n+1} \nn
 &-& \frac{c_1 \zeta_k^2}{2k} \left[
\frac{1+c_2 2 \zeta_k^2 \cos(2\theta_k) (-k\eta)^2 + c_2 2 \zeta_k^2 \cos[2(m+1)\theta_k
+
\pi m]   (-k\eta)^{-2m-2}}{\sin (2\theta_k) (-k\eta)^{2m+2}  + c_2 \zeta_k^2 \sin
(4\theta_k)  (-k\eta)^{2m+4} - c_3 \zeta_k^{-2m-2}  \sin[2m\theta_k + \pi m] } \right]
(-k\eta)^{2m+3}.
\earr
Since in this case $m= -5/2+6\bepsilon$ and $n= 3/2-6\bepsilon$, and considering
only the first dominant term in the expansion around $-k\eta \to 0$, we finally obtain
the expression for the power spectrum presented in \eq{PSescalarcaso4}.

The procedure to obtain the tensor power spectra is analogous to the one outlined in the
previous subsection, but clearly the difference is that $\hat \varTheta_{\nk}^{R,I}
= \hat p_{\nk}^{R,I}$. In the following, we will only present the results.

For case (iii), the tensor power spectrum is given by
\beq
P_t(k) = \frac{2^{2\nu_t+1}H^2 \Gamma^2(\nu_t)}{  M_P^2 \pi^3}
(-k\eta)^{-2\nu_t+1} F_3(\lambda_k,\nu_t), \qquad  \nu_t \equiv \frac{1}{2} + \epsilon,
\eeq
and, consequently, the tensor-to-scalar ratio is $r = 16 \epsilon$, which is the same as
the standard prediction of slow roll inflation.

For case (iv), the formula for the tensor power spectrum is
\beq
P_t(k) = \frac{2}{9 \pi^2}  \left( \int_{-\infty}^\eta
\frac{d\tilde \eta}{z_T^2} \right)^2 \left(\frac{k}{|aH|} \right)^{-2\mu_t+3}
F_4(\lambda_k,\mu_t), \qquad \mu_t \equiv \frac{3}{2} - 6 \bepsilon = \mu_s.
\eeq
Therefore, the tensor-to-scalar ratio is exactly the same as the one shown in \eq{rQMCU}.

\bibliography{bibliografia}
\bibliographystyle{apsrev}

\end{document}